\documentclass[pra,letterpaper]{revtex4}

\usepackage{graphicx}
\usepackage{amsmath}
\usepackage{amssymb}
\usepackage{color}

\begin{document}
\title{Frequency noise and intensity noise of next-generation gravitational-wave detectors with RF/DC readout schemes}
\author{K.~Somiya, Y.~Chen, S.~Kawamura$^\mathrm{A}$, and N.~Mio$^\mathrm{B}$,\\
\small\textit{Max-Planck-Institut f\"ur Gravitationsphysik, Am M\"uhlenberg 1, 14476 Potsdam, Germany},\\
\small\textit{National Astronomical Observatory of Japan, 2-21-1 Osawa Mitaka-shi, Tokyo, 181-8588, Japan}$^\mathrm{A}$,\\
\small\textit{University of Tokyo, 5-1-5 Kashiwanoha Kashiwa-shi, Chiba, 277-8562, Japan}$^\mathrm{B}$}
\date{April 2006}


\begin{abstract}
The sensitivity of next-generation gravitational-wave detectors such as Advanced LIGO and LCGT should be limited mostly by quantum noise with an expected technical progress to reduce seismic noise and thermal noise. Those detectors will employ the optical configuration of resonant-sideband-extraction that can be realized with a signal-recycling mirror added to the Fabry-Perot Michelson interferometer. While this configuration can reduce quantum noise of the detector, it can possibly increase laser frequency noise and intensity noise. The analysis of laser noise in the interferometer with the conventional configuration has been done in several papers, and we shall extend the analysis to the resonant-sideband-extraction configuration with the radiation pressure effect included. We shall also refer to laser noise in the case we employ the so-called DC readout scheme.
\end{abstract}

\maketitle

\section{Introduction}

\begin{figure}[htbp]
\begin{center}
 \includegraphics*[height=6cm]{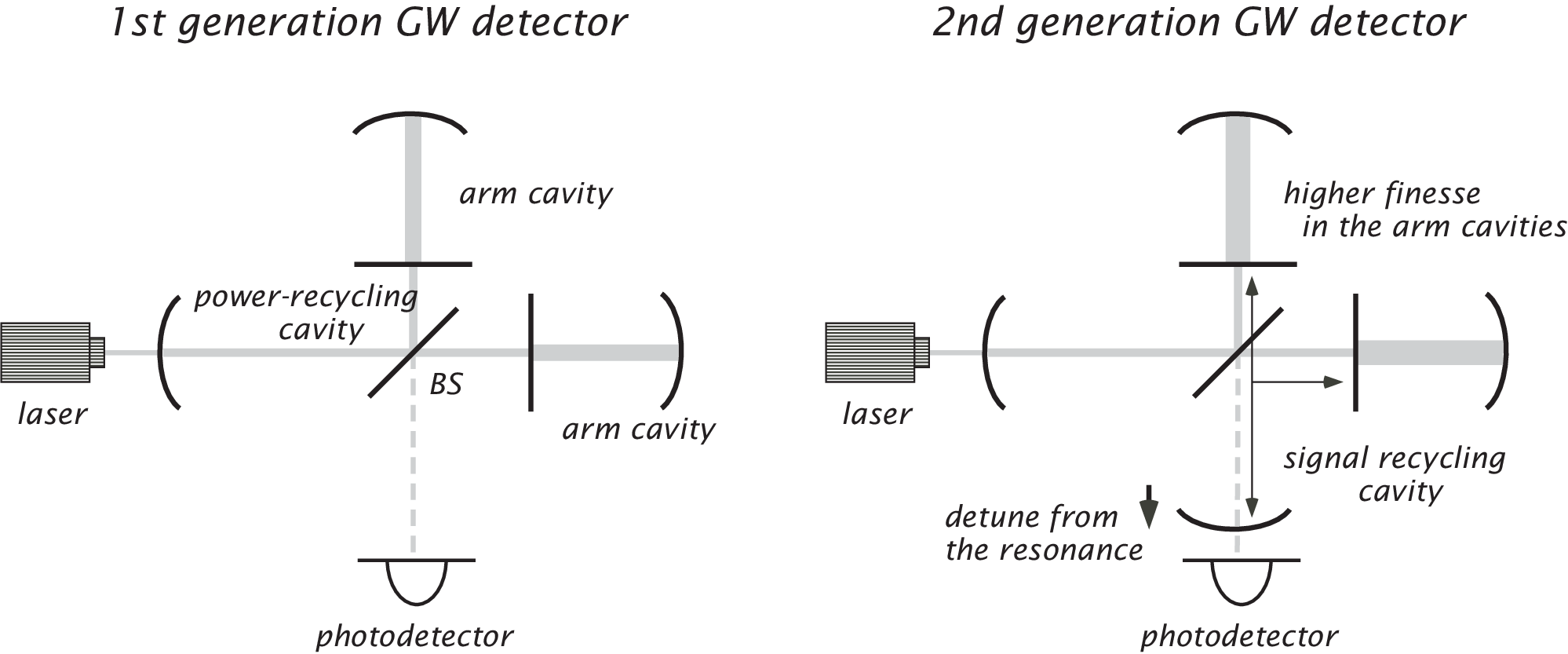}
 \caption{Power-recycled Fabry-Perot Michelson interferometer (left) and power-recycled RSE interferometer (right). The figure is what has been used in reference~\cite{SomiyaAO}.\label{fig:FPMI}}
\end{center}
\end{figure}

Gravitational waves are ripples of space-time curvature, predicted by Einstein's general theory of relativity. Attempts to detect these waves directly have been on-going for several decades, initially with resonant-mass detectors, and recently also with laser interferometers.  Currently, several large-scale ground-based detectors, such as LIGO~\cite{LIGO} in the US, VIRGO~\cite{VIRGO} and GEO600~\cite{GEO} in Europe, and TAMA300~\cite{TAMA} in Japan, have begun operation, and may detect gravitational waves in next few years. LIGO, VIRGO, and TAMA300 detectors are the Michelson interferometer with a Fabry-Perot cavity in their orthogonal arms (left panel of Fig.~\ref{fig:FPMI}).  The Michelson interferometer is controlled to keep the anti-symmetric port dark fringe so that all the differential signal caused by gravitational waves would appear at that port, or the signal extraction port, while the incident beam, together with all its intensity and phase fluctuations, is reflected back toward the laser. The reflected light can be reinjected to the interferometer via a power-recycling mirror located between the laser and the interferometer. In this idealized situation, there will be no laser noise. In reality, however,  inevitable  mismatches between the two arms,  e.g., in optical losses, storage time, length,  etc.,  will couple laser noise into the detection port, and may consequently affect the sensitivity of the interferometer. 

With the operation of first-generation detectors, development of next-generation detectors, with much better sensitivity, such as Advanced LIGO~{\cite{AdLIGO}\cite{AdLIGO2}} in the US and LCGT~\cite{LCGT} in Japan, is underway. In addition to the employment of more advanced seismic isolation systems and higher quality optics, improvement in sensitivity in next-generation detectors is achieved by lowering quantum noise, which is in turn realized by increasing optical power circulating in the arm cavities (lowering shot noise), using heavier test masses (lowering radiation-pressure noise),  and using the optical technique of resonant-sideband-extraction, or RSE~\cite{Mizuno}.  In RSE configurations, arm cavity finesse is much higher than in first-generation detectors. This alone will lower detection bandwidth significantly, but by adding a signal recycling mirror at the detection port  (Fig.~\ref{fig:FPMI}), one can modify the optical response function flexibility. For example, if the signal recycling cavity, i.e., the cavity formed by the signal-recycling mirror and the average position of the two front mirrors, is kept resonant for the carrier light, then the effective reflectivity of the front mirrors decreases for the signal sidebands and one can broaden the detection bandwidth. One can also detune the signal-recycling cavity from resonance, optical sensitivity is then improved around a particular frequency. The  former configuration is often referred to as {\it broadband RSE} and the latter {\it  detuned RSE} (See reference~\cite{SomiyaThesis} for details). In detuned RSE configurations, the signal light coming out from the dark port of the Michelson interferometer, after being reinjected back into the arms with a phase shift, will beat with the carrier light and modulate light amplitude in the arms, and hence radiation-pressure forces exerted on the mirrors. This so-called optical spring effect will improve the sensitivity at the opto-mechanical resonant frequency in addition to the improvement at the optical resonant frequency~\cite{Alessandra}. Advanced LIGO will employ detuned RSE and LCGT will start operation with the broadband RSE configuration with the possibility of changing to detuned RSE in the future.

Laser noise of the power-recycled Fabry-Perot Michelson interferometer has been analyzed by Camp {\textit{et al}} and by Sigg {\textit{et al}} in references~{\cite{Camp}\cite{Sigg}}, and laser noise of the RSE interferometer has been analyzed by Mason in reference~\cite{JimThesis}. These analyses focus on the imperfection of the arms and the filtering effect in the recycling cavities. In 2001, the radiation pressure effect of the detuned RSE interferometer was reported by Buonanno and Chen~\cite{Alessandra}. The signal and quantum fluctuation coming out from the interferometer are reinjected with a phase shift and experience the optical spring effect. Laser noise will experience the same effect, and besides, frequency noise can be converted to intensity noise via an rms fluctuation of the arm cavities, and both that noise and original intensity noise result in radiation pressure noise even if the configuration is broadband RSE. In this paper, we study laser noise of both broadband and detuned RSE configurations with all the radiation pressure effects taken into account. We will consider both the conventional RF modulation-demodulation readout scheme, in which reference light is provided by RF sidebands~\cite{Weiss} and the DC readout scheme, in which reference light is provided by utilizing mismatches between the arms~\cite{DC}. Our results will provide guidance in the design of next-generation interferometers. They are also given in analytic forms, which will help intuitive understanding of laser-noise coupling. 

Our paper is organized as follows: in Sec.~\ref{sec:conv}, we give a brief review of the analysis of laser noise in power-recycled Fabry-Perot Michelson interferometers,  putting it into the quadrature representation, and combining the effect of radiation pressure. In Sec.~\ref{sec:rse}, we go on to study laser noise in RSE configurations, and apply our results to possible plans for Advanced LIGO and LCGT. In Sec.~\ref{sec:summary}, we summarize our main conclusions.

\section{Laser noise of a conventional interferometer}
\label{sec:conv}
Laser noise in low-power Fabry-Perot Michelson interferometers has been studied systematically by Camp et al.~\cite{Camp} and Sigg et al.~\cite{Sigg}. These, combined with classical radiation pressure noise coupled through arm imbalances~\cite{Winkler}, already give us all the necessary results for such configurations. In this section, we collect these results, and put them into the quadrature representation, which is convenient for subsequent analysis of RSE configurations.  

In the following, after discussing signal generation and input laser noise in the quadrature  representation (Sec.~\ref{subsec:sigrf}),  we shall deal separately with all possible channels laser noise can couple into the output of a Fabry-Perot Michelson interferometer, namely: (i) optical coupling from the bright port to the dark port (with mirrors fixed), around carrier frequency (Sec.~\ref{subsec:noisedc}), (ii) optical coupling from the bright port to the dark port (with mirrors fixed), around upper and lower RF modulation frequencies (Sec.~\ref{subsec:noiserf}), (iii) coupling from laser (intensity) noise to mirror motion, and then to the output light at around carrier frequency (Sec.~\ref{subsec:noisemirror}). 

\subsection{Signal generation and input laser noise in the quadrature representation}
\label{subsec:sigrf}

For interferometers with arm lengths much shorter than the wavelength of gravitational waves, the effect of the wave can be regarded solely as tidal forces acting on mirror-endowed test masses. In interferometric gravitational-wave detectors, we measure the differential motions induced by these tidal forces.  

We consider a power-recycled Fabry-Perot Michelson interferometer with input carrier amplitude $E_0 $ and frequency $\omega_0$, power recycling gain (in amplitude),
\begin{equation}
\label{eq:prgain}
g_\mathrm{pr}=\frac{\sqrt{1-r_\mathrm{R}^2}}{1-r_\mathrm{R}r_0}\,,
\end{equation}
where $r_0$ and $r_\mathrm{R}$ are the reflectivity of an arm cavity and the power-recycling mirror, respectively. The carrier, after being amplified by power recycling, split by the beamsplitter, and phase-modulated by a sinusoidal differential mirror motion of $\pm\Re\{ x(\omega) e^{-i\omega t}\}$, will produce the following two reflected fields after each gets reflected again from the beamsplitter, before they are to be recombined:
\begin{eqnarray}
E^{(\pm)}_\mathrm{ref}\simeq\frac{1}{2}g_\mathrm{pr}E_0e^{-i\omega_0t}\left[1\pm\delta(\omega) e^{-i\omega t}\mp\delta^*(\omega) e^{i\omega t}\right]\label{eq:signal}.
\end{eqnarray}
Here the amplitude of the upper and lower sidebands generated by mirror motion is given by
\begin{equation}
\label{eq:delta}
\delta(\omega)=\frac{\omega_0x(\omega)}{c}\frac{N}{1-i\omega/\omega_c}, 
\end{equation}
with
\begin{equation}
N=4/T, \quad \omega_c = T c/(4L)\,.\label{eq:cavitypole}
\end{equation}
The factor $N$ can roughly be regarded as the number of round trips an average photon takes during its life time in the cavity, this gives a signal amplification effect in low frequencies. 
The quantity $\omega_c$, the cavity pole frequency, corresponds to the average photo life time; as we see from Eq.~\eqref{eq:delta}, sensitivity drops when motion frequency $\omega$ is comparable to or larger than $ \omega_c$, because phase modulation can change sign during a photon life time.

In the quadrature representation, instead of using upper and lower modulation sidebands, we consider amplitude and phase modulations. For example, for a carrier frequency of $\omega_0$, modulation frequency of $\omega$, and upper and lower sidebands $a(\pm\omega)$, we would like to have
\begin{equation}
\label{quadconversion}
\Re \left\{e^{- i\omega_0 t} \left[a(\omega) e^{-i\omega t}+a(-\omega) e^{i\omega t}\right]\right\}
= \Re\left[ \sqrt{2} a_1(\omega) e^{-i\omega t}\right] \cos\omega_0 t 
 +\Re\left[\sqrt{2}   a_2(\omega) e^{-i\omega t}\right]\sin\omega_0 t\,.
 \end{equation}
In this way, if we superimpose a reference light of $A \cos\omega_0t$, the fields $a_{1,2}(\omega)$ will then become Fourier components of amplitude and phase modulations.  For this reason, they can be called amplitude and phase quadratures, respectively.   Equation~\eqref{quadconversion} corresponds to the definitions of
\begin{equation}
\label{eq:quadratureconversion}
a_1(\omega) \equiv \frac{a(\omega) + a^*(-\omega)}{\sqrt{2}}\,,\quad a_2(\omega) \equiv \frac{a(\omega) - a^*(-\omega)}{\sqrt{2}i}\,.
\end{equation}
We usually use a 2-dimensional vector, $\mbox{\boldmath $a$}$ to jointly represent the two quadratures, $a_1$ and $a_2$. 

Returning to the response to gravitational-wave-induced motion, if we denote the quadrature-vector representation of the output field as  $\mbox{\boldmath $b$}$, then from Eqs.~\eqref{eq:signal} and~\eqref{eq:quadratureconversion}, we deduce 
\begin{eqnarray}
\label{eq:out:quad}
\left (
\begin{array}{@{\,}c@{\,}}
b_1\\
b_2
\end{array}\right )_\mathrm{sig}=\frac{\sqrt{2}g_\mathrm{pr}E_0\omega_0x}{L(\omega_\mathrm{c}-i\omega)}\left (
\begin{array}{@{\,}c@{\,}}
0\\
1
\end{array}\right ),\label{eq:SG}
\end{eqnarray}
where the top and the bottom are the amplitude and the phase quadrature, respectively. Intuitively, this means that the output signal light corresponds to a phase modulation to the carrier. 

We now turn to the description of the input laser, and its noise, in the quadrature picture. Frequency noise is regarded as phase-modulated sidebands to the carrier light, which is derived in the following way with $\delta\nu$ as frequency fluctuation of the laser:
\begin{eqnarray}
&&\frac{d\phi(t)}{dt}=\omega_0-\int_0^{+\infty}\left(2\pi\delta\nu(\omega)e^{-i\omega t}+2\pi\delta\nu^*(\omega)e^{i\omega t}\right)\frac{d\omega}{2\pi};\ \ \ \ \ \ \delta\nu^*(\omega)=\delta\nu(-\omega)\nonumber\\
&\rightarrow&\ \phi(t)\ =\omega_0t-\int_0^{+\infty}\left(\frac{2\pi\delta\nu(\omega)}{-i\omega}e^{-i\omega t}+\frac{2\pi\delta\nu^*(\omega)}{i\omega}e^{i\omega t}\right)\frac{d\omega}{2\pi}\nonumber\\
&\rightarrow&E_\mathrm{in}(t)=E_0e^{-i\phi}=E_0e^{-i\omega_0t}\left(1+\int_0^{+\infty}\frac{\pi\delta\nu(\omega)}{\omega}e^{-i\omega t}\frac{d\omega}{2\pi}-\int_0^{+\infty}\frac{\pi\delta\nu^*(\omega)}{\omega}e^{i\omega t}\frac{d\omega}{2\pi}\right).\label{eq:freqnoise}
\end{eqnarray}
Intensity noise is amplitude-modulated sidebands to the carrier light, expressed with $P$ and $\delta P$ as the laser intensity and its fluctuation, respectively:
\begin{eqnarray}
&&\left|E_\mathrm{in}(t)\right|^2\hbar\omega_0=P+\int_0^{+\infty}\left(\delta P(\omega)e^{-i\omega t}+\delta P^*(\omega)e^{i\omega t}\right)\frac{d\omega}{2\pi};\ \ \ \ \delta P^*(\omega)=\delta P(-\omega)\nonumber\\
&\rightarrow&E_\mathrm{in}(t)=E_0e^{-i\omega_0t}\left(1+\int_0^{+\infty}\frac{\delta P(\omega)}{4P}e^{-i\omega t}\frac{d\omega}{2\pi}+\int_0^{+\infty}\frac{\delta P^*(\omega)}{4P}e^{i\omega t}\frac{d\omega}{2\pi}\right).\label{eq:intnoise}
\end{eqnarray}
Here, $E_\mathrm{in}$ is the amplitude of the electric field with each fluctuation, and $\hbar$ is the Planck constant. From Eqs.~\eqref{eq:quadratureconversion},~\eqref{eq:freqnoise}, and~\eqref{eq:intnoise}, we have
\begin{equation}
\left(\begin{array}{@{\,}c@{\,}}
l_1 \\
l_2
\end{array}
\right)
 = \frac{E_0}{\sqrt{2}}
\left[
\frac{2\pi\delta\nu}{\omega}
\left(
\begin{array}{@{\,}c@{\,}}
0\\ 
1
\end{array}
\right) 
+
\frac{\delta P}{2P}
\left(
\begin{array}{@{\,}c@{\,}}
1\\ 
0
\end{array}
\right)\right]
\,,\quad \omega \neq 0\,.
\end{equation}
as input laser noise in the quadrature representation.

\subsection{Differential Arm-Cavity Reflectivity: Laser noise around DC}
\label{subsec:noisedc}

In this section, we consider the mirrors as held fixed, i.e., their positions not modified by laser noise, and calculate the transfer function from laser noise around the carrier frequency to the detection port. Without any imperfections between two arm cavities, the asymmetry length $\Delta\ell$ between the distances from the front mirrors to the beamsplitter  would be the only reason for laser noise to leak through the signal extraction port. This asymmetry is used in allowing RF sidebands to propagate to the detection port, and in both Advanced LIGO and LCGT will give negligible effects to laser noise. Instead, we focus on the following mismatches between the two cavities:  $\Delta \epsilon$, mismatch in optical losses, $\Delta T$, mismatch in front-mirror power transmittance, and $\Delta L_{\mathrm{D}}$,  the microscopic mismatch in arm lengths (i.e., the deviation from resonance).  Note that the macroscopic mismatch in arm length (i.e., in multiples of half optical wavelength) is very small, and can be neglected. Fractional mismatch will be less than $\sim10^{-6}$.

The reflectivity of each arm cavity can be expressed in terms of its front-mirror power transmittance $T$, macroscopic length $L$ (integer multiple of half wavelength), microscopic length $\Delta L$ (deviation from resonance), and optical loss $\epsilon$, as a function of sideband frequency $\omega$:
\begin{eqnarray}
r_\mathrm{cav}=\frac{\displaystyle 1-\frac{2}{T}\left(\epsilon\pm\frac{\Delta\epsilon}{2}\right)-s_c\left(1\pm\frac{\Delta\omega_\mathrm{c}}{2\omega_\mathrm{c}}\pm\frac{\omega_0\Delta L}{2\omega L}\right)}{\displaystyle 1+s_c\left(1\pm\frac{\Delta\omega_\mathrm{c}}{2\omega_\mathrm{c}}\pm\frac{\omega_0\Delta L}{2\omega L}\right)}\ ,
\end{eqnarray}
where
\begin{eqnarray}
s_\mathrm{c}\equiv\frac{-i\omega}{\omega_\mathrm{c}}\,.
\end{eqnarray}
We have expanded up to leading order in $T$, $\omega L/c$ and $\omega_0  \Delta L/c$; the cavity pole frequency $\omega_c$ has been given in Eq.~\eqref{eq:cavitypole}. The differential reflectivity, which corresponds to twice the transfer function from the bright port to the dark port, is then
\begin{eqnarray}
\Delta r_\mathrm{cav}(\omega)=\frac{\xi}{1+s_\mathrm{c}}+\frac{2s_\mathrm{c}}{(1+s_\mathrm{c})^2}\frac{\Delta{\cal{F}}}{{\cal{F}}}+\frac{2i}{(1+s_\mathrm{c})^2}\frac{\omega_0 \Delta L_\mathrm{D}}{\omega_\mathrm{c}L} \,, \quad \xi \equiv 2\Delta\epsilon/T\,.
\label{eq:deltar}
\end{eqnarray}
Here, instead of $(\Delta \epsilon, \Delta T, \Delta L_\mathrm{D})$ we have chosen to use $(\xi, \Delta \mathcal{F} /\mathcal{F}, \Delta L_\mathrm{D})$ to characterize mismatches between the arms. Note that $\xi$ describes the {\it contrast defect}, i.e., reflectivity imbalance between the two arms even if the interferometer is locked precisely and has no fluctuation. 

Now in order to obtain laser noise at the detection port, we must also calculate laser noise incident on the beamsplitter.  We obtain, after generalizing Eq.~\eqref{eq:prgain} to non-zero frequencies:
\begin{eqnarray}
\label{eq:tpr}
t_\mathrm{pr}(\omega)=g_\mathrm{pr}\frac{1+s_\mathrm{c}}{1+s_\mathrm{cc}}\,,\quad  s_\mathrm{cc}=-i\omega/\omega_\mathrm{cc}\,,\; \omega_\mathrm{cc}=\frac{1+r_0r_\mathrm{R}}{1+r_\mathrm{R}/r_1}\omega_{\rm c}\,.
\end{eqnarray}
Here $\omega_{\rm cc}$  is the cavity pole of the power-recycled arm cavity, with $r_1$ the reflectivity of the front mirror. Note that $t_{\mathbf{pr}}$ is suppressed significantly from its DC value ($g_\mathrm{pr}$), as $\omega$ exceeds $\omega_{\rm cc}$.  This filtering occurs because the bandwidth of the power-recycling cavity at the carrier frequency is small due to the fact that the power-recycling cavity length is set to the anti-resonant condition for the carrier light and only the light in the bandwidth of the arms will have a $\pi$ phase shift and resonate in the recycling cavity.

Combining Eqs.~\eqref{eq:deltar} and \eqref{eq:tpr}, transfer function from the bright port to the dark port is
\begin{equation}
\mathcal{T}(\omega)=
\frac{1}{2} t_{\rm pr}(\omega) \Delta r_{\rm cav}(\omega) = \frac{g_{\rm pr}}{2(1+s_{\rm cc})}\left[
\xi +\frac{2s_{\rm c}}{1+s_{\rm c}}\frac{\Delta \mathcal{F}}{\mathcal{F}} +\frac{2 i}{1+s_{\rm c}}\frac{\omega_0 \Delta L_\mathrm{D}}{\omega_{\rm c} L }\right]\,.
\end{equation}
According to Appendix~\ref{app:rms} of Ref.~\cite{scaling} [i.e., applying Eqs.~\eqref{eq:quadratureconversion} on the output quadrature fields, using the single sideband transfer function to relate to the input single sidebands, and the applying the inverse of Eqs.~\eqref{eq:quadratureconversion} to relate to the input quadrature fields], the quadrature transfer {\it matrix} {\boldmath{$\mathcal{T}$}} can be written in terms of upper and lower sideband transfer functions $\mathcal{T}(\pm \omega)$, as
\begin{eqnarray}
\mbox{\boldmath{$\mathcal{T}$}}(\omega) &=& 
\frac{1}{2}\left[
\begin{array}{cc}
\mathcal{T}(\omega) + \mathcal{T}^*(-\omega)
& i \mathcal{T}(\omega) - i\mathcal{T}^*(-\omega) \\
-i \mathcal{T}(\omega) + i  \mathcal{T}^*(-\omega)
& \mathcal{T}(\omega) + \mathcal{T}^*(-\omega)
 \end{array}
\right] \,.
\end{eqnarray}
As a consequence, laser noise at the dark port is
\begin{eqnarray}
\label{eq:FPMIcarrierac}
\left (
\begin{array}{@{\,}c@{\,}}
b_1\\
b_2
\end{array}\right )_\mathrm{ca} &=& \mbox{\boldmath{$\mathcal{T}$}}(\omega) \left (
\begin{array}{@{\,}c@{\,}}
l_1\\
l_2
\end{array}\right )
\nonumber \\
&=&
\frac{g_{\rm pr}E_0 }{2 \sqrt{2} (1+s_{\rm cc})}
\left(
\begin{array}{ccc}
\displaystyle \xi +\frac{2s_{\rm c}}{1+s_{\rm c}}\frac{\Delta \mathcal{F}}{\mathcal{F}}  & &  \displaystyle 
 -\frac{2}{1+s_{\rm c}}\frac{\omega_0 \Delta L_\mathrm{D}}{\omega_{\rm c} L } \\ \\
\displaystyle 
 \frac{2}{1+s_{\rm c}}\frac{\omega_0 \Delta L_\mathrm{D}}{\omega_{\rm c} L } &&\displaystyle 
\xi +\frac{2s_{\rm c}}{1+s_{\rm c}}\frac{\Delta \mathcal{F}}{\mathcal{F}} 
 \end{array}
\right)
\left(
\begin{array}{@{\,}c@{\,}}
\displaystyle \frac{\delta P}{2 P}  \\ \\
\displaystyle \frac{2 \pi\delta\nu}{\omega}
\end{array}
\right)
\,.
\end{eqnarray}
In the above equation, diagonal terms in the transfer matrix couples laser intensity noise to output amplitude quadrature, and laser frequency noise to output phase quadrature; non-diagonal terms do the opposite. 
Taking the limit of $\omega\rightarrow 0$ with $\delta P/2P\rightarrow1$ and $\delta\nu\rightarrow0$, we obtain the carrier amplitude at the dark port
\begin{equation}
\label{eq:FPMIcarrierdc}
\left (
\begin{array}{@{\,}c@{\,}}
B_1\\
B_2
\end{array}\right )
=
\frac{g_{\rm pr}E_0 }{2 \sqrt{2} }
\left(
\begin{array}{@{\,}c@{\,}}
\displaystyle \xi \\ \\
\displaystyle 
 \frac{2\omega_0 \Delta L_\mathrm{D}}{\omega_{\rm c} L }
\end{array}
\right)\,.
\end{equation}
Here we use capital $B_{1,2}$ to emphasize the fact that they are amplitudes of sinusoidal fields, instead of Fourier components. 
Below the arm cavity pole frequency (including DC), behavior of the transfer matrix {\boldmath{$\mathcal{T}$}} can be understood very easily in the quadrature picture:  the contrast defect $\xi$ induces differential rescalings of the input carrier vector (including its low-frequency fluctuations), it therefore causes a difference between the reflected vectors that is directly proportional  to the input one;  on the other hand, because cavity detuning causes rotation of output quadratures,  $\Delta L_\mathrm{D}$ induces differential rotations of the input vector, thereby causing a difference between the reflected vectors that is  $90^\circ$ rotated from the input vector.  More explanations are given in Fig.~\ref{fig:phasor}. This is a superposition of quadratures at the carrier and at audio sideband frequencies, which then is equivalent to the way of description with what is called phasor-diagram~\cite{Mizuno}. Other possible fluctuation fields are amplitude and phase fluctuation of an oscillator that generates the RF modulation, which may appear via an imbalance of the upper and the lower RF sidebands~\cite{Camp}, but they are not described in Fig.~\ref{fig:phasor}

\subsection{Laser noise around RF sidebands}
\label{subsec:noiserf}

It is necessary to prepare a reference light at the signal extraction port to obtain the gravitational-wave signal with its phase information. What is usually done in a conventional interferometer is to arrange an electro-optic modulator between the laser and the power-recycling mirror and to prepare a macroscopic difference ($=\Delta\ell$) between distances from the beamsplitter to the two front mirrors, so that RF phase-modulated sidebands appear at the signal extraction port. The amplitude of the RF-sideband field at the signal extraction port is defined to be $\Gamma E_0$ with $E_0$ the input carrier amplitude. In this paper, we only consider RF modulation-demodulation schemes with a single modulation frequency.

\begin{figure}[htbp]
\begin{center}
 \includegraphics*[height=4.5cm]{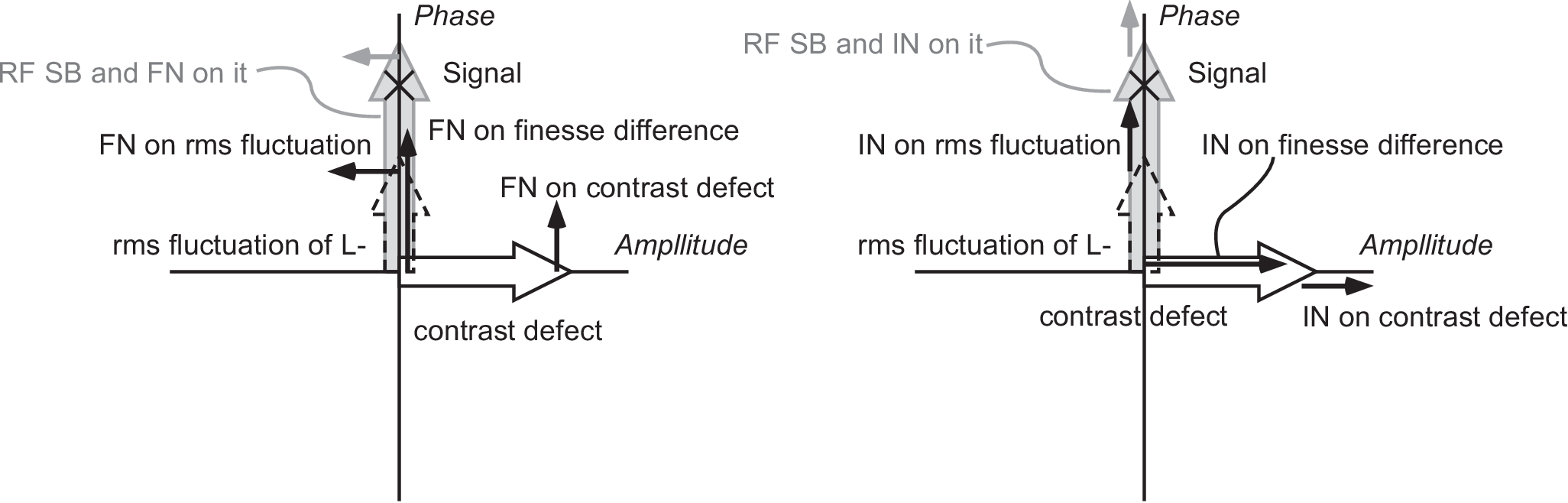}
 \caption{Frequency noise (left) and intensity noise (right) on the phasor diagram with the leaked carrier light at the dark port.\label{fig:phasor}}
\end{center}
\end{figure}

Because RF sidebands were initially generated from the noisy input laser, when they propagate into the detection port, they will carry laser-noise sidebands around RF modulation frequencies (see Fig.~\ref{fig:phasor}).  These laser-noise sidebands will generate laser noise in the output, by beating with carrier light coming out from the detection port. As we demonstrate in Appendix~\ref{app:RF}, under certain reasonable assumptions, we can describe the effect of an entire modulation-demodulation scheme using two vectors, one at DC and the other in the frequency band of gravitational waves,
\begin{equation}
\left (
\begin{array}{@{\,}c@{\,}}
B_1\\
B_2
\end{array}\right )_\mathrm{RF}
 = 
\Gamma E_0  \left (
\begin{array}{@{\,}c@{\,}}
\sin\zeta\\
\cos\zeta
\end{array}\right ) \,,\quad 
\left (
\begin{array}{@{\,}c@{\,}}
b_1\\
b_2
\end{array}\right )_\mathrm{RF}=\Gamma E_0
\left[
\frac{2\pi\delta\nu}{\omega}\left (
\begin{array}{@{\,}c@{\,}}
-\cos{\zeta}\\
\sin{\zeta}
\end{array}\right )
+\frac{\delta P}{2P}\left (
\begin{array}{@{\,}c@{\,}}
\sin{\zeta}\\
\cos{\zeta}
\end{array}\right )\right]\,,\label{eq:RFSB}
\end{equation}
in the sense that the signal of the modulation-demodulation scheme will be
\begin{equation}
\label{SGRF}
\mathrm{SG}_{\rm RF}=
\left (
\begin{array}{@{\,}c@{\,}}
B_1\\
B_2
\end{array}\right )_\mathrm{RF}\bullet
\left (
\begin{array}{@{\,}c@{\,}}
b_1\\
b_2
\end{array}\right )_\mathrm{sig}\,,
\end{equation}
while laser noise of the scheme will be 
\begin{equation}
\label{LNRF}
\mathrm{LN}_{\rm RF}=
\left (
\begin{array}{@{\,}c@{\,}}
B_1\\
B_2
\end{array}\right )_\mathrm{RF}\bullet
\left (
\begin{array}{@{\,}c@{\,}}
b_1\\
b_2
\end{array}\right )_\mathrm{ca}
+ 
\left (
\begin{array}{@{\,}c@{\,}}
B_1\\
B_2
\end{array}\right )_\mathrm{ca}\bullet
\left (
\begin{array}{@{\,}c@{\,}}
b_1\\
b_2
\end{array}\right )_\mathrm{RF}\,.
\end{equation}
Here $\bullet$ means the operation to take an inner product of two vectors. It must be emphasized here that the fields $\mathbf{B}_{\rm RF}$ and $\mathbf{b}_{\rm RF}$ are {\it not} quadrature fields directly describing RF sideband fields or their laser-noise sidebands, which oscillate at around $\omega_\mathrm{m}$. Instead, these effective fields were obtained taking into account also the demodulation procedure. In some sense, they are down-conversions of RF fields into the DC region, via demodulation (see Appendix~\ref{app:RF}). 

In Fabry-Perot Michelson interferometers, we have $\zeta=0$, and 
\begin{equation}
\label{RFFPMI}
\left (
\begin{array}{@{\,}c@{\,}}
B_1\\
B_2
\end{array}\right )_\mathrm{RF}
 = 
\Gamma E_0  \left (
\begin{array}{@{\,}c@{\,}}
0\\
1
\end{array}\right ) \,,\quad 
\left (
\begin{array}{@{\,}c@{\,}}
b_1\\
b_2
\end{array}\right )_\mathrm{RF}
 = 
\Gamma E_0  
\left[
-\frac{2\pi i \delta\nu}{\omega}
\left(
\begin{array}{@{\,}c@{\,}}
1\\ 
0
\end{array}
\right)
+
\frac{\delta P}{2P}
\left(
\begin{array}{@{\,}c@{\,}}
0\\ 
1
\end{array}
\right)\right] \,.
\end{equation}
 The RF sidebands are fixed in the phase quadrature in the case of the balanced configuration but not in a detuned RSE~\cite{SomiyaPRD}, which will be calculated in Sec.~\ref{sec:rse}. The RF sidebands experience a filter of the power-recycling cavity but the cavity pole of the filter is higher than the observation frequencies so that the effect is negligible.

Inserting Eqs.~\eqref{eq:FPMIcarrierac}, \eqref{eq:FPMIcarrierdc}, and \eqref{RFFPMI} into Eq.~\eqref{LNRF}, we obtain laser frequency noise and intensity noise:
\begin{eqnarray}
{\rm FN}_\mathrm{conv}
&=&\Gamma g_\mathrm{pr} E_0^2\frac{\pi}{\sqrt{2}\omega}\left\{\left(\frac{1}{1+s_\mathrm{cc}}-1\right)\xi+\frac{2s_\mathrm{c}}{(1+s_\mathrm{cc})(1+s_\mathrm{c})}\frac{\Delta{\cal{F}}}{{\cal{F}}}\right\}\delta\nu, \nonumber\\
&=&\Gamma g_\mathrm{pr} E_0^2\frac{\pi}{\sqrt{2}\omega}\left\{\frac{-s_\mathrm{cc}}{1+s_\mathrm{cc}}\xi+\frac{2s_\mathrm{c}}{(1+s_\mathrm{cc})(1+s_\mathrm{c})}\frac{\Delta{\cal{F}}}{{\cal{F}}}\right\}\delta\nu, \\
{\rm IN}_\mathrm{conv}&=&
\Gamma g_\mathrm{pr} E_0^2\frac{1}{2\sqrt{2}}\frac{\omega_0 \Delta L_\mathrm{D}}{\omega_\mathrm{c}L}\left\{\frac{1}{(1+s_\mathrm{cc})(1+s_\mathrm{c})}+1\right\} \frac{\delta P}{P}\,.
\end{eqnarray}
On the other hand, the signal output obtained from Eqs.~(\ref{eq:SG}) and (\ref{SGRF}) is
\begin{eqnarray}
\mathrm{SG}_\mathrm{conv}&=&\Gamma g_\mathrm{pr} E_0^2\frac{\sqrt{2}\omega_0x}{L\omega_\mathrm{c}(1+s_\mathrm{c})}.
\end{eqnarray}
Signal-referred frequency noise and intensity noise are  then
\begin{eqnarray}
h_\mathrm{conv}^{\rm FN}&=&\frac{\pi}{2\omega_0}\left\{\frac{-s_\mathrm{cc}(1+s_\mathrm{c})}{s_\mathrm{c}(1+s_\mathrm{cc})}\ \xi+\frac{2}{1+s_\mathrm{cc}}\frac{\Delta{\cal{F}}}{{\cal{F}}}\right\} \delta \nu,\label{FNTFFPMI}\\
h_\mathrm{conv}^{\rm IN}&=&\frac{1+(1+s_\mathrm{c})(1+s_\mathrm{cc})}{1+s_\mathrm{cc}}\frac{\Delta L_\mathrm{D}}{4L}  \frac{\delta P}{P}\,.\label{INTFFPMI}
\end{eqnarray}

\begin{figure}[htbp]
\begin{center}
 \includegraphics*[width=12cm]{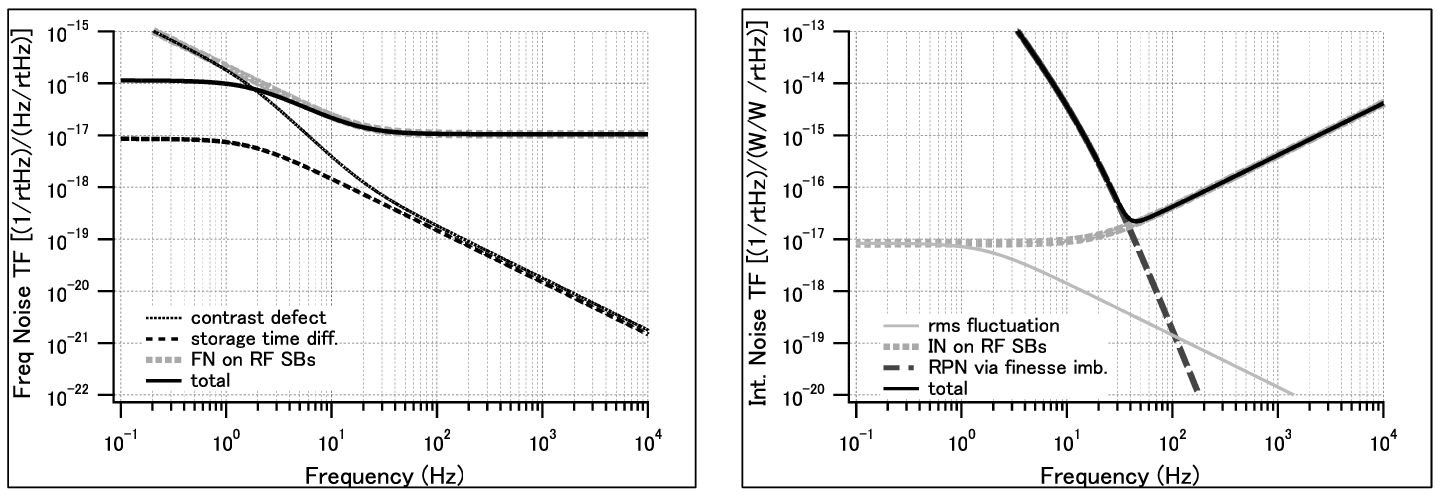}
 \caption{Frequency noise (left) and intensity noise (right) of the power-recycled Fabry-Perot Michelson interferometer, calculated. Here the transfer function from each fluctuation to the equivalent strain sensitivity are shown.\label{fig:IgorPR}}
\end{center}
\end{figure}
Note that here we assume the RF readout scheme, and the DC readout scheme will be explained later in Sec.~\ref{sec:DC}. Figure~\ref{fig:IgorPR} (left) shows the calculated result. Frequency noise on the carrier and on the RF sidebands are cancelled each other at the low frequencies, and that on the RF sidebands are dominant at the high frequencies. The parameters used here are $L=3$~km, ${\cal{F}}=1250$, and the power-recycling gain $g_\mathrm{pr}^2\simeq10$, which are the anticipated parameters for the LCGT interferometer without its signal-recycling mirror. The same parameters with the signal-recycling mirror will be used later. Contrast defect of 26~mW at the dark port is assumed, which is derived from the differential cavity reflectivity of $\xi=0.012$ or the round-trip loss difference of $\Delta\epsilon=30$~ppm. In the case of AdLIGO without the signal-recycling mirror, contrast defect would be 55~mW with $\xi=0.012$. The contrast defect in the current detectors is higher than this value but can be realized with the installation or the improvement of an output mode-cleaner, which is used to filter out junk light that makes contrast defect large in the current detectors. The finesse difference of $0.5~\%$ is assumed, which is also challenging compared with $1\sim2~\%$ that is the value of the current detectors. As for the rms fluctuation, $\Delta L_\mathrm{D}=10^{-13}$~m is assumed.

\subsection{Laser noise coupled through radiation pressure imbalance}
\label{subsec:noisemirror}

The results shown in Eqs.~\eqref{FNTFFPMI} and \eqref{INTFFPMI} coincide to the results in references~{\cite{Camp}\cite{Sigg}}, and we are going to extend the analysis to the RSE configurations. Before the extension to the RSE, however, we shall take into account laser noise through radiation pressure imbalances, which would appear in next-generation interferometers with a high-power laser. 

The radiation pressure force is proportional to fluctuation of the power in the cavity:
\begin{eqnarray}
F_\mathrm{rp}=\frac{2\Delta W_\mathrm{arm}}{c},
\end{eqnarray}
where the power fluctuation in each arm is
\begin{eqnarray}
\Delta W_\mathrm{arm}^\mathrm{(i)}&=&\frac{1}{1+s_\mathrm{cc}}t_\mathrm{BS}^2\frac{4\sqrt{2}I_\mathrm{BS}}{T}\frac{\delta P}{4P}\ \ \ \mathrm{(inline\ arm)}\ ,\label{eq:Winl}\\
\Delta W_\mathrm{arm}^\mathrm{(p)}&=&\frac{1}{1+s_\mathrm{cc}}r_\mathrm{BS}^2\frac{4\sqrt{2}I_\mathrm{BS}}{T}\frac{\delta P}{4P}\ \ \ \mathrm{(perpendicular\ arm)}\ .\label{eq:Wper}
\end{eqnarray}
Here $I_\mathrm{BS}$ is the power at the beamsplitter, $r_\mathrm{BS}$ and $t_\mathrm{BS}$ are the reflectivity and the transmittance of the beamsplitter, respectively. The motion of the mirror in differential mode of the arm cavities is given by $X_\mathrm{rp}=-4\Delta W_\mathrm{arm}/mc\omega^2$ with the consideration of effective mass $\mu=m/2$ in each cavity, where $m$ is the mass of each mirror. At the dark port appears intensity noise due to the differential motion of two arm cavities only if the parameters are unbalanced. Adding the power-recycling cavity and assuming the imbalance $\sigma=r_\mathrm{BS}^2-t_\mathrm{BS}^2$ in the beamsplitter, the mass difference $m\pm\Delta m/2$, and the cavity pole difference $\omega_\mathrm{c}\pm\Delta\omega_\mathrm{c}/2$, which is equivallent to the finesse difference $\mathcal{F}\mp\Delta\mathcal{F}/2$, between Eqs~\eqref{eq:Winl} and \eqref{eq:Wper}, we have
\begin{eqnarray}
b_\mathrm{imb}=\frac{1}{(1+s_\mathrm{c})(1+s_\mathrm{cc})}\frac{\sqrt{2} g_\mathrm{pr}E_0 I_\mathrm{BS}\omega_0}{m\omega^2L^2\omega_\mathrm{c}^2}\frac{\delta P}{P}\left[\sigma+\frac{\Delta\omega_\mathrm{c}/\omega_\mathrm{c}}{(1+s_\mathrm{c})^2}+\frac{\Delta m}{2m}\right]\,.
\end{eqnarray}
in the phase quadrature at the dark port. We can simplify the output vector as
\begin{eqnarray}
\mbox{\boldmath $b$}_\mathrm{imb}=\frac{t_\mathrm{pr}E_0}{\sqrt{2}}\left(\frac{\delta P}{4P}\right)\kappa e^{2i\beta}\left[\sigma+\frac{\Delta\omega_\mathrm{c}/\omega_\mathrm{c}}{(1+s_\mathrm{c})^2}+\frac{\Delta m}{2m}\right]
\left (
\begin{array}{@{\,}c@{\,}}
0\\
1
\end{array}\right ),\label{eq:CRPNviaImb}
\end{eqnarray}
with $\kappa$ as the coupling coefficient of the radiation pressure effect, and $\beta$ as the phase shift of the audio sidebands in the arm cavity:
\begin{eqnarray}
\kappa\ \ &=&\frac{I_\mathrm{BS}/I_\mathrm{SQL}\cdot 2\omega_\mathrm{c}^4}{\omega^2(\omega^2+\omega_\mathrm{c}^2)},\label{eq:kappa}\\
I_\mathrm{SQL}&=&\ \frac{mL^2\omega_\mathrm{c}^4}{4\omega_0},\\
\beta\ \ &=&\arctan{\left(\frac{\omega}{\omega_\mathrm{c}}\right)}.
\end{eqnarray}
These parameters, which have been introduced in reference~\cite{KLMTV}, will be used in the extension to the RSE system. The meaning of $I_\mathrm{SQL}$ is the light power, with which quantum radiation pressure noise will be as large as the shot noise level at the cavity pole frequency. Here we neglect the influence of a pendulum that is used to make a suspended mirror nearly free-mass beyond the resonant frequency ($\sim1$~Hz).

Figure~\ref{fig:IgorPR} (right) shows the calculated result of intensity noise, where $0.5~\%$ of the cavity-pole imbalance is included. Classical radiation pressure noise via the imbalance is the largest at low frequencies.

\section{Laser noise in RSE interferometers}
\label{sec:rse}

As is reported in reference~\cite{Alessandra}, in {\it detuned RSE} configurations, signal light reflected by the signal-recycling mirror and reinjected back into the interferometer will beat with the incident beam, modulate the radiation-pressure force acting on the test masses, thereby modifying test-mass dynamics. Quantum-noise-limited gravitational-wave sensitivity is significantly influenced by this feedback mechanism. In particular, in addition to the optical resonance due to detuning, there can also exist another resonance, at a lower frequency, around which gravitational-wave sensitivity is also enhanced. Laser noise in RSE configurations will also experience the feedback effect due to signal recycling.  As we shall see in this section, in {\it both} broadband {\it and } detuned RSE configurations, intensity noise due to contrast defect and frequency noise due to arm-length imbalance will both beat with carrier light in the arm cavities and cause classical radiation pressure noise. 

Besides analyzing features of laser noise coupling, we also study the so-called DC readout schemes, in which the local oscillator is provided by (partially intentional) arm imbalances. In these schemes, on the one hand, arm imbalance must be big enough, in order to provide a strong enough local oscillator, on the other hand, it must be small enough such that laser noise remain suppressed compared to more fundamental noise sources, e.g., quantum noise. 

\subsection{AC laser noise coupling}

We shall use the input-output relation of the RSE interferometer described in reference~\cite{Alessandra}, which consists of the propagation of quantum fluctuation and that of the signal in the phase quadrature. Here we can concentrate on the classical part of the equation that includes laser noise. Besides, the propagation of the signal, or classical noise, in the amplitude quadrature can also be derived in the same way, which is shown in reference~\cite{Somiya40m}. The output of a conventional power-recycled Fabry-Perot Michelson interferometer is given by
\begin{eqnarray}
\label{inout1}
\left( \begin{array}{@{\,}c@{\,}} b_1 \\ b_2 \end{array}\right)_\mathrm{ca+sig} = {e^{2i\beta} \left(
\begin{array}{cc}
1 & 0  \\
- \kappa & 1
\end{array}\right) \left( \begin{array}{@{\,}c@{\,}} d_1 \\ d_2 \end{array}\right)}
 + \frac{\sqrt{2} g_{\rm pr} E_0 \omega_0 x}{L(\omega_{\rm c}-i\omega)} 
\left( \begin{array}{@{\,}c@{\,}} 0 \\ 1 \end{array}\right)
 + \mbox{\boldmath{$\mathcal{T}$}} \left( \begin{array}{@{\,}c@{\,}} \ell_1 \\ \ell_2 \end{array}\right) 
 + \left( \begin{array}{@{\,}c@{\,}} 0 \\ b_\mathrm{imb} \end{array}\right).
\end{eqnarray}
\begin{figure}[htbp]
\begin{center}
 \includegraphics*[width=5cm]{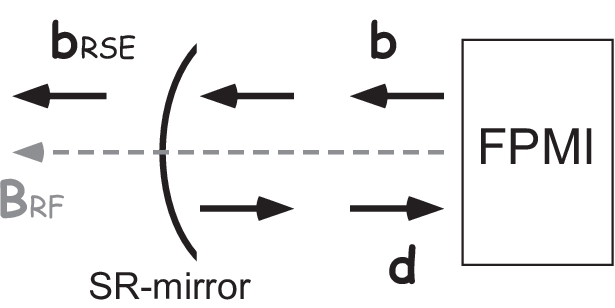}
 \caption{Input-output relation of laser noise in the signal-recycling cavity.\label{fig:inputoutput}}
\end{center}
\end{figure}
The first term of the right member expresses the field entering to the interferometer from its dark port (see Fig.~\ref{fig:inputoutput}). The matrix in that term shows the coupling of a component in the amplitude quadrature to the phase quadrature through the radiation pressure effect. Here a possible imbalance of $\kappa$ in the two arm cavities is neglected. With the signal-recycling mirror, the signal and laser noise that appear at the dark port of the Michelson interferometer are reinjected to the interferometer:
\begin{eqnarray}
 \label{inout2}
\left( \begin{array}{@{\,}c@{\,}} d_1 \\ d_2 \end{array}\right) = \rho 
\left(
\begin{array}{cc}
\cos2 \phi & -\sin2\phi \\
\sin2\phi & \cos 2\phi 
\end{array}
\right)
\left( \begin{array}{@{\,}c@{\,}} b_1 \\ b_2 \end{array}\right)_\mathrm{ca+sig}.
\end{eqnarray}
Here $\rho$ is the amplitude reflectivity of the signal recycling mirror, and $\phi$ detuning phase of the signal recycling cavity from anti-resonance ($\phi=\pi/2$ for the broadband RSE). Then the signal and laser noise leak out the RSE interferometer:
\begin{eqnarray}
\label{inout3}
\left( \begin{array}{@{\,}c@{\,}} b_1 \\ b_2 \end{array}\right)_{\rm RSE} =  \tau
\left(
\begin{array}{cc}
\cos \phi & -\sin\phi \\
\sin\phi & \cos \phi 
\end{array}
\right)\left( \begin{array}{@{\,}c@{\,}} b_1 \\ b_2 \end{array}\right)_\mathrm{ca+sig},
\end{eqnarray}
where $\tau$ is the amplitude transmittance of the signal recycling mirror. For the moment, we only use Eq.~\eqref{inout1} for non-zero, i.e., AC, frequencies, for which laser noise will be evaluated. Subtleties arise  DC, because there must be pendulum restoring force and control forces that balance the DC radiation pressure from the carrier. For AC, we really mean frequencies above $\sim 1$\,Hz, and the slow drifting behavior of $\Delta L_\mathrm{D}$, as first mentioned in Sec.~\ref{subsec:noisedc} and treated in Appendix.~\ref{app:rms}, takes place below this frequency. As discussed in Appendix~\ref{app:rms}, we use $(\Delta L_\mathrm{D})_{\rm rms}$ for calculating AC laser noise spectrum.  We postpone discussion of DC light to Sec.~\ref{subsec:dclight}. 

Solving Eqs.~\eqref{inout1} - \eqref{inout3}, we obtain the AC coupling of laser noise
\begin{eqnarray}
\left (
\begin{array}{@{\,}c@{\,}}
b_1\\
b_2
\end{array}\right )_{\rm RSE}
=\frac{\tau g_\mathrm{pr}E_0 }{M}
\left (
\begin{array}{@{\,}cc@{\,}}
D_{11}&D_{12}\\
D_{21}&D_{22}
\end{array}\right )
\left[
\left(\begin{array}{@{\,}c@{\,}}
a_1\\
a_2
\end{array}\right) 
+
\frac{\sqrt{2}\omega_0 x}{L(\omega_{\rm c}-i\omega)}
\left (
\begin{array}{@{\,}c@{\,}}
0 \\1
\end{array}\right )
\right]
\ ,\label{eq:RSE}
\end{eqnarray}
where
\begin{eqnarray}
&&M=1+\rho^2e^{4i\beta}-2\rho e^{2i\beta}\left(\cos{2\phi}+\frac{\kappa}{2}\sin{2\phi}\right),\nonumber\\
\nonumber\\
&&D_{11}=(1-\rho e^{2i\beta})\cos{\phi}-\rho e^{2i\beta}\kappa\sin{\phi}\ ,\nonumber\\
&&D_{12}=-(1+\rho e^{2i\beta})\sin{\phi}\ ,\nonumber\\
&&D_{21}=(1+\rho e^{2i\beta})\sin{\phi}-\rho e^{2i\beta}\kappa\cos{\phi}\ ,\nonumber\\
&&D_{22}=(1-\rho e^{2i\beta})\cos{\phi}\ .
\end{eqnarray}
Here the input vector $\mbox{\boldmath{$a$}}=(\mbox{\boldmath{$\mathcal{T}$}}\mbox{\boldmath{$l$}}+\mbox{\boldmath{$b$}}_\mathrm{imb})/g_\mathrm{pr}E_0$ indicates the input laser noise field around the carrier frequency in a conventional interferometer, normalized by the input field. We shall write down the components again:
\begin{eqnarray}
a_1&=&-\frac{\pi\delta\nu}{\sqrt{2}\omega}\frac{2}{(1+s_\mathrm{cc})(1+s_\mathrm{c})}\frac{\omega_0 \Delta L_\mathrm{D}}{\omega_\mathrm{c}L}+\frac{\delta P}{4\sqrt{2}P}\left\{\frac{\xi}{1+s_\mathrm{cc}}+\frac{2s_\mathrm{c}}{(1+s_\mathrm{cc})(1+s_\mathrm{c})}\frac{\Delta{\cal{F}}}{{\cal{F}}}\right\}\label{eq:a1}\\
a_2&=&\frac{\pi\delta\nu}{\sqrt{2}\omega}\left\{\frac{\xi}{1+s_\mathrm{cc}}+\frac{2s_\mathrm{c}}{(1+s_\mathrm{cc})(1+s_\mathrm{c})}\frac{\Delta{\cal{F}}}{{\cal{F}}}\right\}\nonumber\\
&&\ \ \ \ \ +\frac{\delta P}{4\sqrt{2}P}\frac{1}{(1+s_\mathrm{cc})(1+s_\mathrm{c})}\left\{\frac{2\omega_0 \Delta L_\mathrm{D}}{\omega_\mathrm{c}L}+\kappa(1+s_\mathrm{c})^2\left[\sigma+\frac{\Delta\omega_\mathrm{c}/\omega_\mathrm{c}}{(1+s_\mathrm{c})^2}+\frac{\Delta m}{2m}\right]\right\}.\label{eq:a2}
\end{eqnarray}

As discussed in Appendix~\ref{app:RF}, RF sidebands and laser noise around them will be downconverted by the demodulation process, and provide an effective local oscillator and effective noise contributions. In the most general case, one has, at the detection port:
\begin{eqnarray}
\left (
\begin{array}{@{\,}c@{\,}}
B_1\\
B_2
\end{array}\right )_\mathrm{RF}&=&\Gamma E_0
\left (
\begin{array}{@{\,}c@{\,}}
\sin{\zeta}\\
\cos{\zeta}
\end{array}\right )\ ,\label{eq:RFRSEdc}\\
\left (
\begin{array}{@{\,}c@{\,}}
b_1\\
b_2
\end{array}\right )_\mathrm{RF}&=&\Gamma E_0
\left[
\frac{2\pi\delta\nu}{\omega}\left (
\begin{array}{@{\,}c@{\,}}
-\cos{\zeta}\\
\sin{\zeta}
\end{array}\right )
+\frac{\delta P}{2P}\left (
\begin{array}{@{\,}c@{\,}}
\sin{\zeta}\\
\cos{\zeta}
\end{array}\right )\right]\,.\label{eq:RFRSEac}
\end{eqnarray}
These are the same equations as is written in Eq.~\eqref{eq:RFSB}. In broadband RSE configuration planned for LCGT,  the upper and the lower RF sidebands are balanced, and one can only detect the output phase quadrature, with $\zeta = \pi/2$. In most detuned RSE configurations, the sidebands are unbalanced (i.e., one of them much stronger than the other), and can be regarded as a single sideband. The quadrature of the effective RF local oscillator $\zeta$ is equal to the demodulation phase, which can assume any value. 

The readout quadrature, or readout phase, represented by $\zeta$ above, plays an important role in the next-generation gravitational-wave detectors that accommodate a high-power laser. As is shown in references~\cite{Alessandra}\cite{SomiyaPRD}\cite{KLMTV}\cite{Miyakawa40m}, quantum noise in any configuration and the signal in detuned configurations change the frequency response with the readout phase $\zeta$, which can be chosen to optimize the signal-to-noise ratio to gravitational-wave signals. The variation can be seen either with RF readout or DC readout. In detuned RSE configurations, where we can tune the readout phase, it is desirable that laser noise not limit the sensitivity with any readout phase.

\subsection{DC light at the detection port}
\label{subsec:dclight}

In order to study laser noise, and to understand the DC readout scheme, the output DC light leaking to the detection port must be calculated. This involves the initial arm-length mismatch, the slow drift of arm lengths, arm length offset introduced by control system, and the feedback of light by the signal-recycling mirror.  Here for DC, we really mean frequencies below 1~Hz, in which the slow drift of arm cavity lengths is dramatic (see Appendix.~\ref{app:rms}).  

Equation~\eqref{inout1} should be modified by taking the limit of $\omega \rightarrow 0$. One can see there is a term with $\kappa$, which becomes infinity by taking the limit of $\omega \rightarrow 0$ (see Eq.\eqref{eq:kappa}). In reality, however, because of pendulum restoring force, $\omega^2$ in the denominator should be replaced by $\omega^2-\omega_p^2$, where $\omega_p^2$ is the pendulum frequency. Actual quantity $\kappa_0$ shall be prepared (In Advanced LIGO, $\kappa_0$ can be $\sim 10^4$--$10^6$). 

Now the DC components can be divided to three parts. One is the term that has $\kappa_0$ in its numerator. This is a radiation pressure offset induced by contrast defect. The rest part of contrast defect is the second one, which stays there even if $\kappa=0$. It leaks to the detection port with a phase shift due to the detuning of the signal-recycling cavity. The other is the term that is proportional to $\Delta\tilde{L}_\mathrm{D}$. Equation~\eqref{eq:RSE} becomes ($\omega\rightarrow0$; $\delta P/2P\rightarrow1$, $\delta\nu\rightarrow0$, $b_\mathrm{imb}\rightarrow0$, and $x\rightarrow0$)
\begin{eqnarray}
\label{RSEcarrierout1}
\left (
\begin{array}{@{\,}c@{\,}}
B_1\\
B_2
\end{array}\right )_{\rm RSE}=
\frac{g_\mathrm{pr}E_0}{2\sqrt{2}}
\left (
\begin{array}{@{\,}c@{\,}}
\displaystyle\frac{\mathstrut\tau\xi(1-\rho)\cos{\phi}}{M_{\rm free}}-\frac{\tau(1+\rho)\sin{\phi}}{M_{\rm dc}}\frac{2\omega_0(\Delta\tilde L_\mathrm{D} + \Delta\tilde L_\mathrm{cd})}{L\omega_\mathrm{c}}\\

\displaystyle\frac{\mathstrut\tau\xi(1+\rho)\sin{\phi}}{M_{\rm free}}+\frac{\tau(1-\rho)\cos{\phi}}{M_{\rm dc}}\frac{2\omega_0(\Delta\tilde L_\mathrm{D} + \Delta\tilde L_\mathrm{cd})}{L\omega_\mathrm{c}}
\end{array}\right )\ ,
\end{eqnarray}
where 
\begin{eqnarray}
M_{\rm free}&=&M\ (\omega=0,\ \kappa=0)=1+\rho^2-2\rho\cos{2\phi}\ ,\\
M_{\rm dc}&=&M\ (\omega=0)=1+\rho^2-2\rho\cos{2\phi}-\kappa_0\rho\sin{2\phi}\ ,
\end{eqnarray}
and we have defined
\begin{eqnarray}
\label{Lcd}
\Delta \tilde L_\mathrm{cd}=- \frac{L\omega_\mathrm{c}}{2\omega_0}\frac{\xi\kappa_0\rho(\cos{2\phi}-\rho)}{M_{\rm free}}\,
\end{eqnarray}
as the offset of the arm cavities from their resonance in the differential mode caused by radiation pressure of contrast defect reinjected through the signal-recycling mirror. This offset should be cancelled out by feedback control with an electric offset, which appears in $\Delta  \tilde L_\mathrm{ctrl}$. Note that $\Delta \tilde L_\mathrm{ctrl}$ is not only the control offset to cancel radiation pressure force and also includes the offset to drive the mirrors to the operation points from their initial suspension points. Even with feedback controls, there still remains a microscopic mismatch in the arm length that we have also seen in the calculation in Sec.~\ref{subsec:noisedc} as the rms fluctuation. Here it is written by $\Delta \tilde L_\mathrm{rms}$.

In the calculations above, $\Delta \tilde L_\mathrm{cd}$, $\Delta \tilde L_\mathrm{ctrl}$, and $\Delta \tilde L_\mathrm{rms}$ are written in their {\it free} displacements of the differential mode before feedback from the optical spring. One can see a suppression factor in the second terms of Eq.~\eqref{RSEcarrierout1}, which means the offset displacements in the differential mode are suppressed by the spring. It also means that, in the case we want to add the control offset to cancel the radiation pressure or to choose a quadrature of the DC readout scheme, we need to impose big force to move rather small displacement that is actually seen.

In fact, it is the {\it actual} displacement that we need for calculating laser noise coupling. We shall define the total actual displacement in the differential mode:
\begin{eqnarray}
\label{RSEtotalmotion}
\Delta  L_\mathrm{D}  &=& \Delta L_\mathrm{ctrl} +\Delta   L_\mathrm{rms}+\Delta  L_\mathrm{cd}\nonumber\\
&=& \frac{M_{\rm free}}{M_{\rm dc}} \left(\Delta\tilde L_\mathrm{ctrl} + \Delta\tilde L_\mathrm{rms} + \Delta\tilde L_\mathrm{cd}\right)\,.
\end{eqnarray}
Equation~\eqref{RSEcarrierout1} becomes
\begin{eqnarray}
\left (
\begin{array}{@{\,}c@{\,}}
B_1\\
B_2
\end{array}\right )_{\rm RSE}=
\frac{g_\mathrm{pr}E_0}{2\sqrt{2}}
\left (
\begin{array}{@{\,}c@{\,}}
\displaystyle\frac{\mathstrut\tau\xi(1-\rho)\cos{\phi}}{M_{\rm free}}-\frac{\tau(1+\rho)\sin{\phi}}{M_{\rm free}}\frac{2\omega_0\Delta   L_\mathrm{D}}{L\omega_\mathrm{c}}\\
\displaystyle\frac{\mathstrut\tau\xi(1+\rho)\sin{\phi}}{M_{\rm free}}+\frac{\tau(1-\rho)\cos{\phi}}{M_{\rm free}}\frac{2\omega_0 \Delta  L_\mathrm{D}}{L\omega_\mathrm{c}}
\end{array}\right )\ .\label{eq:RSEdc}
\end{eqnarray}

When $\Delta L_\mathrm{rms}=\Delta L_\mathrm{ctrl}=0$, i.e., the arm lengths are originally perfectly matched in absence of RSE, and there is no control force applied,  Eqs.~\eqref{RSEcarrierout1} and \eqref{Lcd} give,
\begin{equation}
\label{eq:RSEDCnocon}
\left (
\begin{array}{@{\,}c@{\,}}
B_1\\
B_2
\end{array}\right )_{\rm RSE}=
\frac{\tau g_{\rm pr} E_0 \xi}{2\sqrt{2} M_{\rm dc}}
\left[
\begin{array}{@{\,}c@{\,}}
(1-\rho)\cos\phi-\kappa_0\rho\sin\phi\\
(1+\rho)\sin\phi-\kappa_0\rho\cos\phi\end{array}\right]\,.
\end{equation}
For broadband RSE ($\phi=\pi/2$), Eq.~\eqref{eq:RSEDCnocon} with the approximation $\kappa_0 \gg 1$ becomes
\begin{eqnarray}
\label{RSEcarrierbroadband}
\left (
\begin{array}{@{\,}c@{\,}}
B_1\\
B_2
\end{array}\right )_{\rm RSE}&=&
g_\mathrm{pr}E_0\frac{\tau\xi}{2\sqrt{2}(1+\rho)^2}
\left (
\begin{array}{@{\,}c@{\,}}
-\kappa_0\rho\\
1+\rho
\end{array}\right )\quad \mathrm{(broadband\ RSE)}
\end{eqnarray}
In this situation, contrast defect light drives significant amount of mirror motion, due to the absence of optical-spring suppression. Result from Eq.~\eqref{RSEcarrierbroadband} can often become unphysical, if $\kappa_0 \tau\xi$ becomes larger than unity, because there should not be more light emerging from the detection port than input on the beamsplitter, and our perturbative approach fails. A control system must be applied to cancel that effect.  For detuned RSE configurations, with $\phi$ not too close to $\pi/2$ (more specifically, $(\phi-\pi/2 )\gg 1/\kappa_0$), Eq.~\eqref{eq:RSEDCnocon} becomes
\begin{eqnarray}
\left (
\begin{array}{@{\,}c@{\,}}
B_1\\
B_2
\end{array}\right )_{\rm RSE}&=&
g_\mathrm{pr}E_0\frac{\tau\xi}{2\sqrt{2}\sin{2\phi}}
\left (
\begin{array}{@{\,}c@{\,}}
\sin{\phi}\\
\cos{\phi}
\end{array}\right )\quad \mathrm{(detuned\ RSE)}.\label{eq:DCwoCTRL}
\end{eqnarray}
In this situation, the resulting mirror motion due to fed-back contrast-defect light is suppressed by optical spring, yet if $\phi$ is too close to $\pi/2$, the output light might still be a significant fraction of all the light incident on the beamsplitter.

\begin{figure}[htbp]
\begin{center}
 \includegraphics*[height=3.5cm]{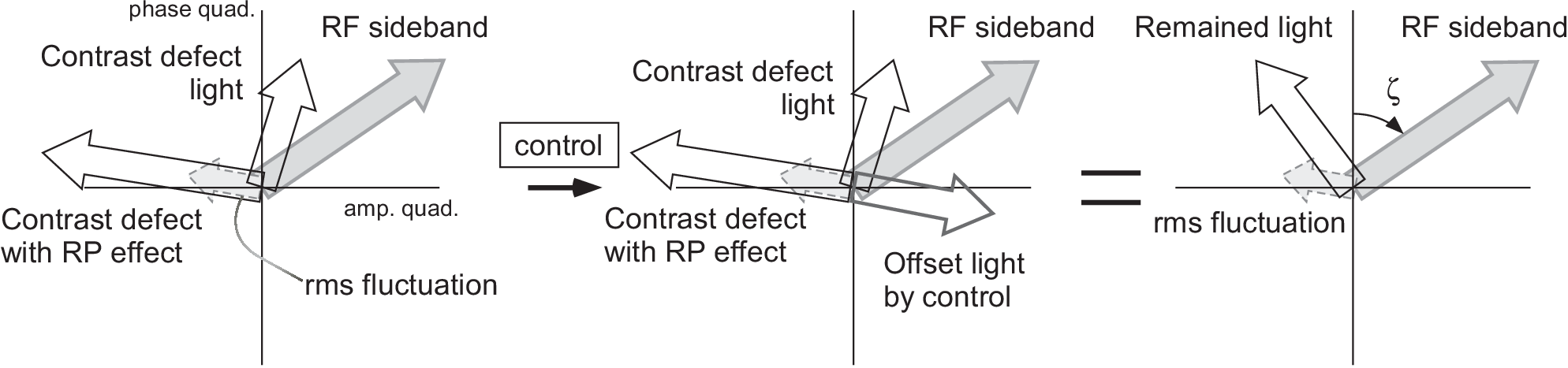}
 \caption{DC components of the carrier light and the RF sidebands at the dark port of the RSE interferometer. The offset light caused by radiation pressure of the reinjected contrast defect light can be suppressed by the control system.\label{fig:offset}}
\end{center}
\end{figure}

A control force must be applied to the differential mode, to allow us to tailor the quadrature at which the carrier emerges at the detection port. As is shown in Fig.~\ref{fig:offset}, when the feedback control with the readout phase $\zeta$ works, a part of the offset caused by radiation pressure that corresponds to the same quadrature to the RF sidebands can be suppressed and the rest part on the other quadrature remains still with rms fluctuation of the control. The remained carrier light due to the contrast defect and its radiation pressure offset should be given by
\begin{eqnarray}
\left (
\begin{array}{@{\,}c@{\,}}
B_1\\
B_2
\end{array}\right )_\mathrm{RSE-rem}=
\frac{g_\mathrm{pr}E_0}{2\sqrt{2}}
\left (
\begin{array}{@{\,}c@{\,}}
\tau\xi'\cos{\zeta}\\
-\tau\xi'\sin{\zeta}
\end{array}\right )\ ,\label{eq:RSEdc2}
\end{eqnarray}
which quadrature is orthogonal to that of RF sidebands in a readout phase of $\zeta$:
\begin{eqnarray}
\left (
\begin{array}{@{\,}c@{\,}}
B_1\\
B_2
\end{array}\right )_\mathrm{RF}=
\Gamma E_0
\left (
\begin{array}{@{\,}c@{\,}}
\sin{\zeta}\\
\cos{\zeta}
\end{array}\right )\,.\nonumber
\end{eqnarray}
To realize this situation, the free displacement due to the control should be
\begin{eqnarray}
\Delta \tilde{L}_\mathrm{ctrl}=-\Delta\tilde{L}_\mathrm{cd}-\frac{M_{\rm dc}}{M_{\rm free}}\frac{\sin{(\phi+\zeta)}+\rho\sin{(\phi-\zeta)}}{\cos{(\phi+\zeta)}-\rho\cos{(\phi-\zeta)}}\frac{L\omega_\mathrm{c}\xi}{2\omega_0}\ ,
\end{eqnarray}
which corresponds to an actual displacement of
\begin{eqnarray}
\Delta L_\mathrm{ctrl}=\left[\frac{\kappa_0\rho(\cos{2\phi}-\rho)}{M_\mathrm{free}}-\frac{\sin{(\phi+\zeta)}+\rho\sin{(\phi-\zeta)}}{\cos{(\phi+\zeta)}-\rho\cos{(\phi-\zeta)}}\right]\frac{L\omega_\mathrm{c}\xi}{2\omega_0}\ .\label{eq:RSEoffset}
\end{eqnarray}
Now the remained carrier light has a modified contrast defect amplitude $\xi'$ instead of $\xi$;
\begin{eqnarray}
\xi'=\frac{\xi}{\cos{(\phi-\zeta)}-\rho\cos{(\phi+\zeta)}}\,.\label{eq:xidash}
\end{eqnarray}
In our notation, any quadrature is achievable, in principle, except: 
\begin{eqnarray}
\label{zetaforbidden}
\zeta^{\rm forbidden}=\arctan{\left(-\frac{1-\rho}{1+\rho}\cot{\phi}\right)}\ .\label{eq:particularphase}
\end{eqnarray}
As $\zeta$ approaches $\zeta^{\rm forbidden}$, amplitude of the output DC light becomes very high, which will increase the contribution of laser noise around RF sidebands.

\begin{figure}[htbp]
\begin{center}
 \includegraphics*[width=12cm]{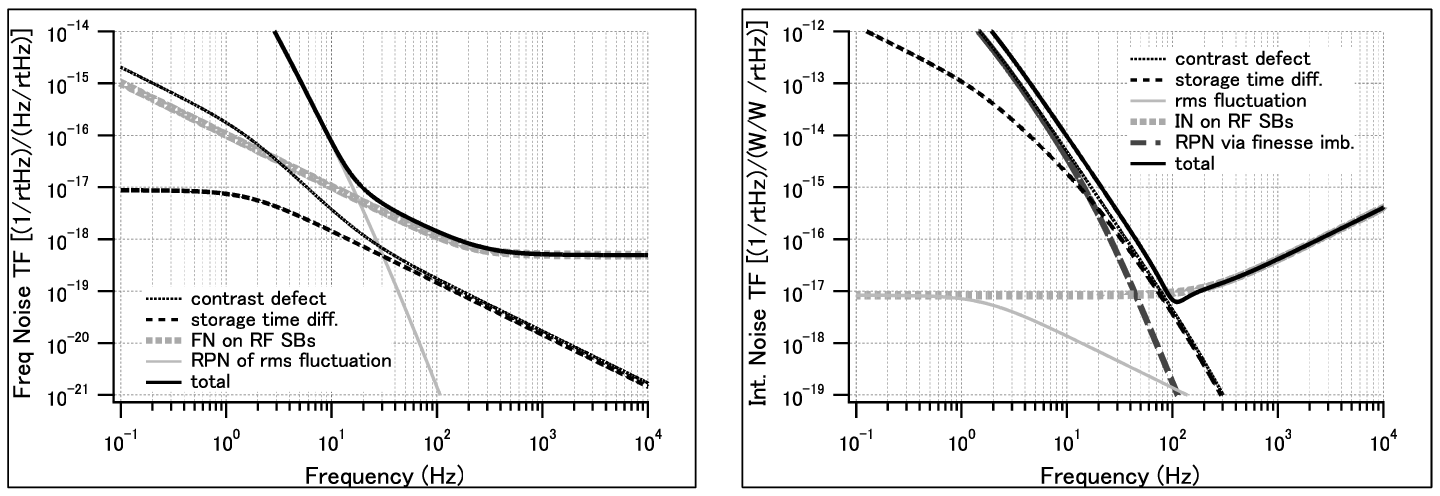}
 \includegraphics*[width=12cm]{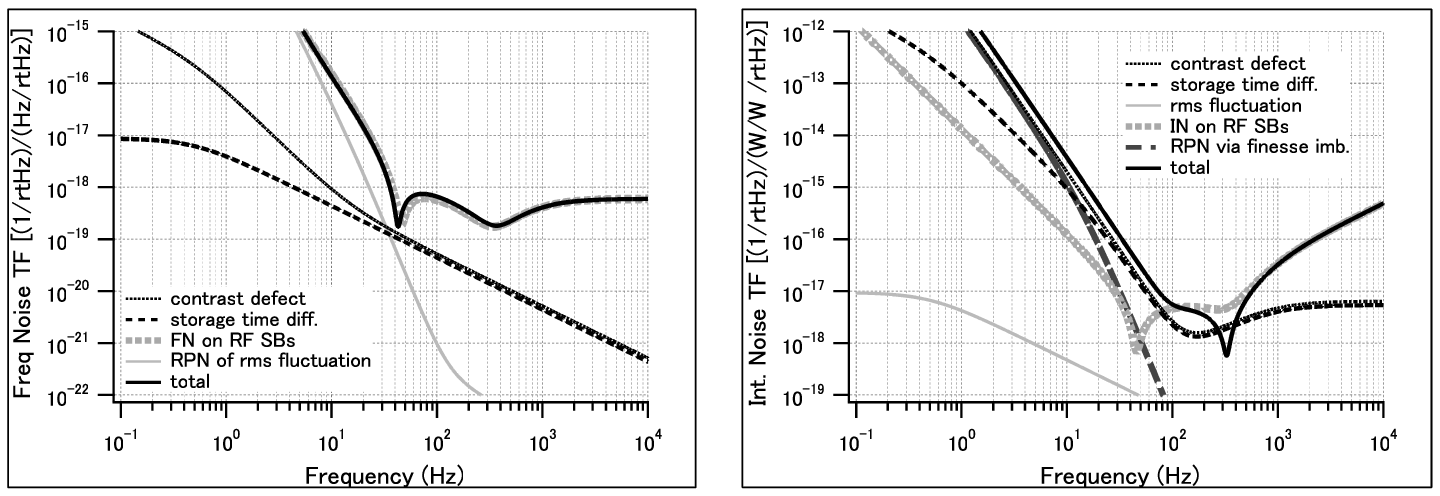}
 \caption{Laser noise of the broadband RSE (top) and the detuned RSE (bottom).\label{fig:finalRF}}
\end{center}
\end{figure}

Finally, the additional carrier that emerges due to $\Delta \tilde  L_\mathrm{rms}$ is given by
\begin{equation}
\left(
\begin{array}{@{\,}c@{\,}}
B_1\\
B_2
\end{array}\right)_{\rm RSE-rms}
=
\frac{\tau g_{\rm pr}E_0}{2\sqrt{2} M_{\rm dc}}\frac{2\omega_0 \Delta \tilde L_\mathrm{rms}}{\omega_c L}\left[
\begin{array}{c}
-(1+\rho)\sin\phi \\
(1-\rho)\cos\phi
\end{array}
\right]\,.\label{eq:RSErms}
\end{equation}

\subsection{RF readout for RSE configurations}

Laser noise with RF readout in the RSE configuration is obtained from the sum of the beat of effective RF local oscillator in Eq.~\eqref{eq:RFRSEdc} with laser noise around DC in Eq.~\eqref{eq:RSE}, and beat of DC light in Eqs.~\eqref{eq:RSEdc2} and \eqref{eq:RSErms} with effective RF laser noise in Eq.~\eqref{eq:RFRSEac}, that is,
\begin{eqnarray}
\left(\mathrm{SG}+\mathrm{LN}\right)_{\rm RSE-RF}=\left(
\begin{array}{@{\,}c@{\,}}
B_1\\
B_2
\end{array}\right)_{\rm RF}
\bullet\left(
\begin{array}{@{\,}c@{\,}}
b_1\\
b_2
\end{array}\right)_{\rm RSE}
+\left[\left(
\begin{array}{@{\,}c@{\,}}
B_1\\
B_2
\end{array}\right)_{\rm RSE-rem}
+\left(
\begin{array}{@{\,}c@{\,}}
B_1\\
B_2
\end{array}\right)_{\rm RSE-rms}\right]
\bullet\left(
\begin{array}{@{\,}c@{\,}}
b_1\\
b_2
\end{array}\right)_{\rm RF}\ ,\label{eq:LNRSERF}
\end{eqnarray}
where the signal is included. Signal-referred laser noise is obtained by taking the ratio from the noise component to the signal in Eq.~\eqref{eq:LNRSERF}. The result is
\begin{eqnarray}
h_{\rm rse}&=&\frac{\omega_\mathrm{c}-i\omega}{\sqrt{2}\omega_0}\frac{(D_{11}a_1+D_{12}a_2)\sin{\zeta}+(D_{21}a_1+D_{22}a_2)\cos{\zeta})}{D_{12}\sin{\zeta}+D_{22}\cos{\zeta}}\nonumber\\
&+&\frac{\omega_\mathrm{c}-i\omega}{\sqrt{2}\omega_0}\frac{M}{D_{12}\sin{\zeta}+D_{22}\cos{\zeta}}\left\{-\frac{\pi\delta\nu}{2\sqrt{2}\omega}\xi'+\frac{1}{M_{\rm dc}}\frac{\sqrt{2}\omega_0\Delta\tilde{L}_\mathrm{rms}}{\omega_\mathrm{c}L}
\left(
\begin{array}{@{\,}c@{\,}}
\sin{(\phi+\zeta)}+\rho\sin{(\phi-\zeta)}\\
\cos{(\phi+\zeta)}-\rho\cos{(\phi-\zeta)}
\end{array}\right)\bullet
\left(
\begin{array}{@{\,}c@{\,}}
\pi\delta\nu/\omega\\
\delta P/4P
\end{array}\right)\right\}\,.\nonumber\\
\label{eq:hrseRF}
\end{eqnarray}
Numerical results with RF readout are shown in Fig.~\ref{fig:finalRF}. The parameters for the broadband RSE are $L=3$~km, $m=30$~kg, $\rho=0.82$, $g_\mathrm{pr}^2=10$, and $\zeta=\pi/2$, which are the parameters for LCGT~\cite{LCGT2}, and the parameters for the detuned RSE are $L=4$~km, $m=40$~kg, $\rho=0.96$, $g_\mathrm{pr}^2=16$, $\phi=\pi/2-0.04$, and $\zeta=\pi/2-0.04$, which are the parameters for Advanced LIGO ($\phi$ and $\zeta$ are not decided yet, so just an example). 

\subsection{DC readout scheme}\label{sec:DC}

Since unbalanced RF sidebands increase the contribution of phase noise of the RF oscillator, the conventional readout scheme with RF demodulation is not good with a detuned configuration. Advanced LIGO will use the so-called DC readout scheme,  in which the local oscillator is not RF sidebands but carrier light leaked to the dark port due to arm imbalances~\cite{DC}. This scheme is useful even with initial LIGO interferometers, since there are many practical advantages~\cite{Rana} and it is the easiest way to remove additional quantum noise at the demodulation process~\cite{AlessandraNSSN}. Based on results given in the previous sections, we study laser noise with the DC readout scheme.

\begin{figure}[htbp]
\begin{center}
 \includegraphics*[height=4.5cm]{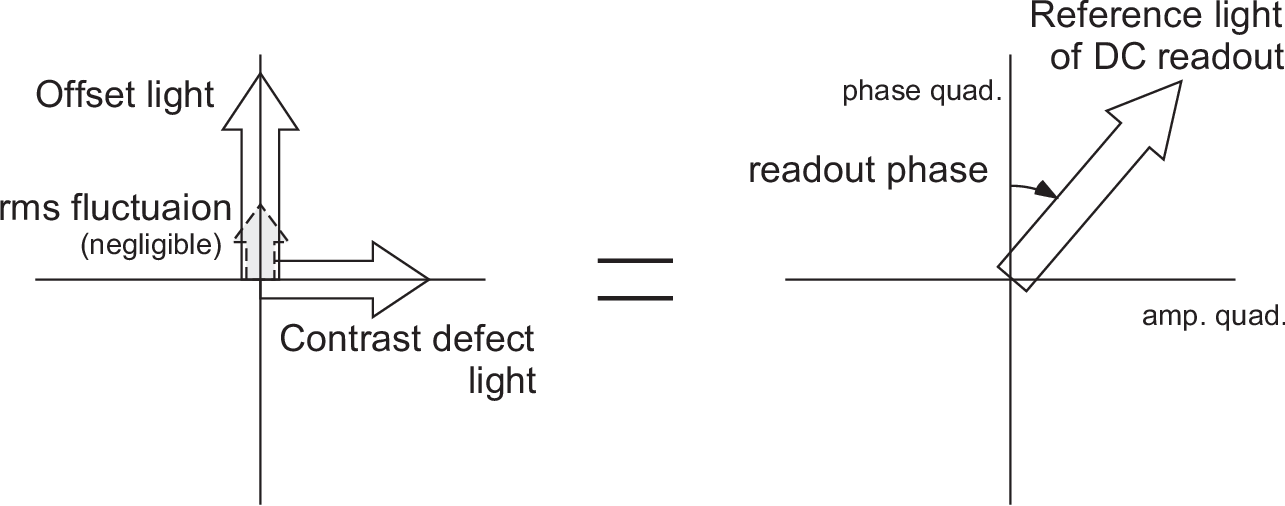}
 \caption{Readout quadrature is determined by the ratio of offset light and contrast defect light.\label{fig:DC}}
\end{center}
\end{figure}

\subsubsection{DC readout for conventional configurations}

As is shown in Eq.~(\ref{eq:FPMIcarrierdc}), for conventional configurations, DC light appears not only in the amplitude quadrature due to contrast defect, but also in the phase quadrature due to differential mismatch in arm lengths, $\Delta L_\mathrm{D}$ (Fig.~\ref{fig:DC}). Also it is possible with DC readout to decrease quantum radiation pressure noise as the price for paying for the sensitivity at high frequencies, if one chooses a quadrature different from the phase quadrature~\cite{KLMTV}. In  currently operating power-recycled Fabry-Perot Michelson interferometers, however, the readout quadrature should be the phase quadrature since the sensitivity is not limited by radiation pressure noise at low frequencies. The offset length $\Delta L_\mathrm{D}$ should meet
\begin{eqnarray}
\Delta L_\mathrm{D}\gg\frac{\xi\omega_\mathrm{c}}{\omega_0}L
\end{eqnarray}
to make the readout quadrature close to the phase quadrature. With the LCGT parameters without the signal-recycling mirror, for example, $\Delta L_\mathrm{D}$ needs to be at least $10^{-10}\sim 10^{-11}$~m, which is larger than its rms fluctuation.

\begin{figure}[htbp]
\begin{center}
 \includegraphics*[width=12cm]{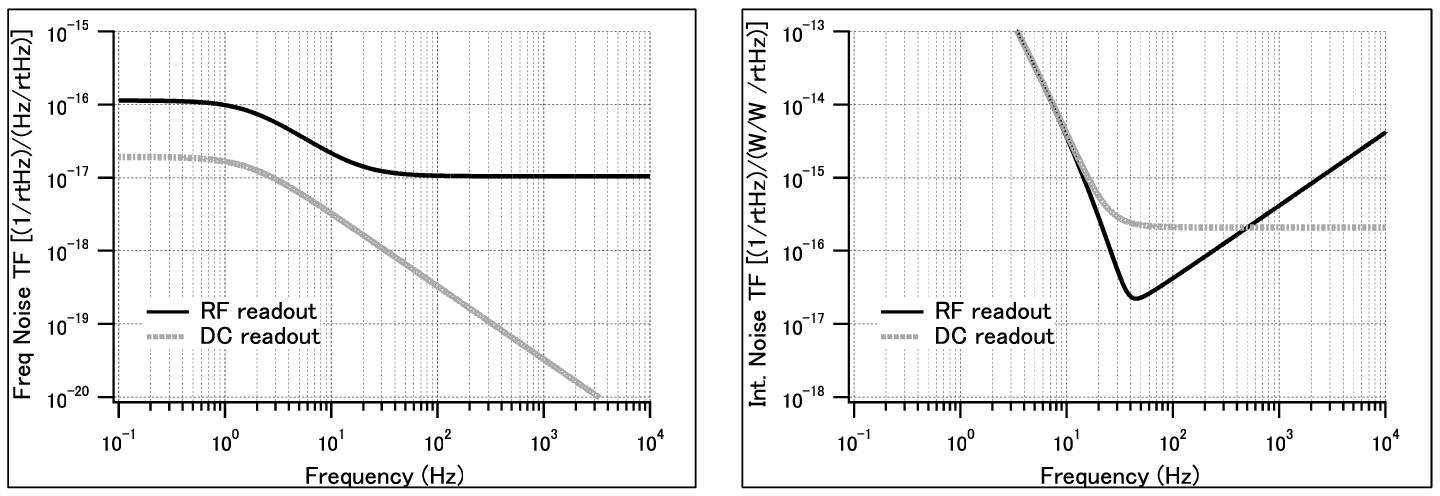}
 \caption{Laser noise with the DC and the RF readout scheme.\label{fig:IgorPRDC}}
\end{center}
\end{figure}

While intensity noise may increase due to the raise of $\Delta L_\mathrm{D}$, frequency noise on the reference light is filtered out at high frequencies unlike that on the RF sidebands and total frequency noise is suppressed. From eqs.~(\ref{eq:SG}), (\ref{eq:FPMIcarrierac}), and (\ref{eq:FPMIcarrierdc}), frequency noise and the signal with DC readout is derived:
\begin{eqnarray}
{\rm FN}_\mathrm{conv-dc}&=&g_\mathrm{pr}^2 E_0^2\frac{\pi\delta\nu}{\sqrt{2}\omega}\frac{2\omega_0\Delta L_\mathrm{D}}{\omega_\mathrm{c}L}\left\{\left[\frac{1}{1+s_\mathrm{cc}}-\frac{1}{(1+s_\mathrm{c})(1+s_\mathrm{cc})}\right]\xi+\frac{2s_\mathrm{c}}{(1+s_\mathrm{cc})(1+s_\mathrm{c})}\frac{\Delta{\cal{F}}}{{\cal{F}}}\right\}\nonumber\\
&=&g_\mathrm{pr}^2 E_0^2\frac{\pi\delta\nu}{\sqrt{2}\omega}\frac{\omega_0\Delta L_\mathrm{D}}{\omega_\mathrm{c}L}\frac{2s_\mathrm{c}}{(1+s_\mathrm{cc})(1+s_\mathrm{c})}\left\{\xi+\frac{2\Delta{\cal{F}}}{{\cal{F}}}\right\},\\
{\rm SG}_\mathrm{conv-dc}&=&g_\mathrm{pr}^2 E_0^2\frac{2\omega_0\Delta L_\mathrm{D}}{\omega_\mathrm{c}L}\frac{\sqrt{2}\omega_0x}{L\omega_\mathrm{c}(1+s_\mathrm{c})}.\label{eq:SGDC}
\end{eqnarray}
Therefore the signal-referred frequency noise level is 
\begin{eqnarray}
h_\mathrm{conv-dc}^{\rm FN}=\frac{\pi}{2\omega_0}\frac{1}{1+s_\mathrm{cc}}\left\{\xi+\frac{2\Delta{\cal{F}}}{{\cal{F}}}\right\} \delta\nu.
\end{eqnarray}
Intensity noise is derived as well:
\begin{eqnarray}
{\rm IN}_\mathrm{conv-dc}&=&g_\mathrm{pr}^2 E_0^2\frac{\delta P}{4\sqrt{2}P}\frac{1}{1+s_\mathrm{cc}}\left\{\left(\frac{\omega_0 \Delta L_\mathrm{D}}{\omega_\mathrm{c}L}\right)^2\frac{4}{1+s_\mathrm{c}}+\xi\left[\xi+\frac{2s_\mathrm{c}}{1+s_\mathrm{c}}\frac{\Delta{\cal{F}}}{{\cal{F}}}\right]\right.\nonumber\\
&&\hspace{3.5cm}\left.+2\kappa e^{2i\beta}(1+s_\mathrm{c})\left[\sigma+\frac{\Delta\omega_\mathrm{c}/\omega_\mathrm{c}}{(1+s_\mathrm{c})^2}+\frac{\Delta m}{2m}\right]\right\},\label{eq:INDC}
\end{eqnarray}
and the signal-referred intensity noise level is given by  the ratio of Eqs.~(\ref{eq:INDC}) and (\ref{eq:SGDC}). The calculated results are shown in Fig.~\ref{fig:IgorPRDC}. Both frequency noise and intensity noise decrease at high frequencies and a difference of frequency noise at low frequencies depends on the loss difference and the arm imbalances.

\subsubsection{DC readout for RSE configurations.}

We have discussed in detail in the previous section the DC light emerging from the detection port, in particular the fact that the quadrature $\zeta$ of DC output light can be tailored by the control system. It was shown that $\zeta$ can take any value, except $\zeta^{\rm forbidden}$ [Cf.~Eq.~\eqref{zetaforbidden}]. In the case of broadband RSE, changing the readout phase from $\pi/2$ makes it possible to remove quantum radiation pressure noise at particular frequency as well as in the conventional configuration~{\cite{SomiyaThesis}\cite{homodyne}}, but here we shall assume the readout phase of $\pi/2$ for simplicity. The DC light at the dark port of the broadband RSE interferometer with additional offset $\Delta L_\mathrm{D}(\sim10^{-11}~\mathrm{m})$ is obtained from Eq.~(\ref{eq:RSEdc}):
\begin{eqnarray}
\left (
\begin{array}{@{\,}c@{\,}}
B_1\\
B_2
\end{array}\right )_\mathrm{BRSE}=
g_\mathrm{pr}E_0\frac{\tau}{1+\rho}
\left (
\begin{array}{@{\,}c@{\,}}
-2\omega_0\Delta L_\mathrm{D}/\omega_\mathrm{c} L\\
\xi
\end{array}\right ),
\end{eqnarray}
and signal-referred laser noise is
\begin{eqnarray}
h_\mathrm{brse-dc}=\left(\frac{D_{11}a_1+D_{12}a_2}{D_{12}}+\frac{D_{21}a_1}{D_{12}}\frac{\xi\omega_\mathrm{c} L}{2\omega_0\Delta L_\mathrm{D}}\right)\frac{\omega_{\rm c} -i\omega}{\sqrt{2}{\omega_0}}\ .
\end{eqnarray}
The first term is laser noise coupled with the offset light and the second term is laser noise coupled with contrast defect. Here we assume that the low-frequency drift of $\Delta L_\mathrm{D}$ due to the rms fluctuation component can be negligible. Transfer function of frequency fluctuation to the dark port is given by $h_\mathrm{n}/\delta\nu$ with $\delta P\rightarrow 0$ and that of intensity fluctuation is given by $h_\mathrm{n}/(\delta P/P)$ with $\delta\nu\rightarrow 0$.

\begin{figure}[htbp]
\begin{center}
 \includegraphics*[width=15cm]{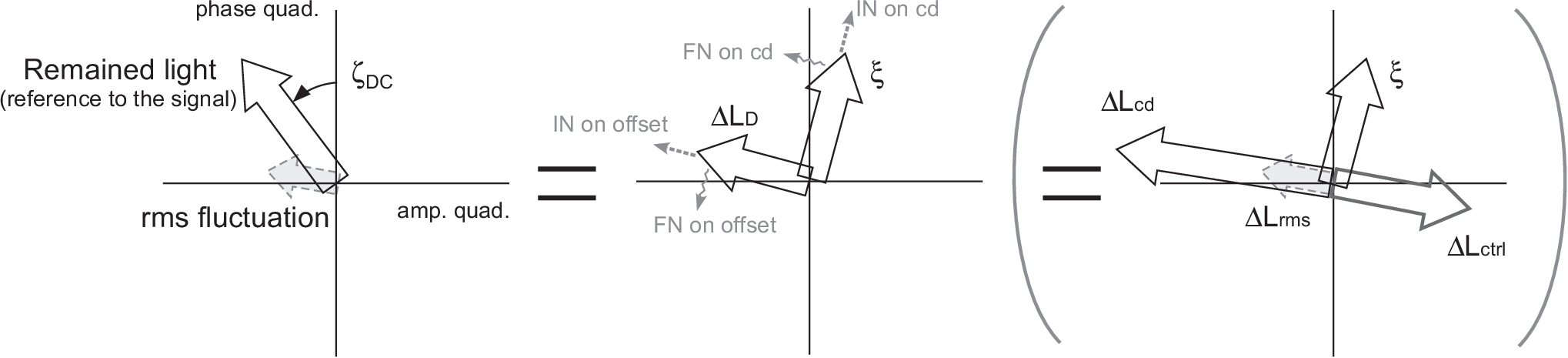}
 \caption{Phasor diagram of the light field at the detection port with the DC readout scheme. The reference light to probe the signal is the sum of contrast defect and the offset light excluding rms fluctuation (left panel). Laser noise is obtained from the calculation of the couplings between dc components and ac components of contrast defect and the offset light including rms fluctuation (middle panel). The offset light consists of rms fluctuation, radiation pressure offset due to contrast defect, and the control light (right panel). Here rms fluctuation should be so small as not to change the readout phase. \label{fig:RSEdcreadout}}
\end{center}
\end{figure}

In the case of detuned RSE with DC readout, it is known that the sensitivity with the readout phase ($\neq\zeta$) being $\pi/2$ is better at low frequencies but worse at the frequency of the best sensitivity than that with the readout phase being 0~{\cite{SomiyaThesis}\cite{Alessandra}}. The radiation pressure offset can be completely removed, with a sufficient control gain, by choosing $\zeta=\phi(\sim\pi/2)$, in which case the arm length mismatch only contains the drift, after suppression from the optical spring: $\Delta L_\mathrm{D} =\Delta L_\mathrm{rms}\sim10^{-13}$~m. If we choose $\zeta$ to be far from $\phi$, for example $\zeta=0$, then the actual mismatch  $\Delta L_\mathrm{D}$ is larger, because an artificial imbalance needs to be introduced to create a local oscillator. For example, 
 $\zeta=0$ and $\phi=\pi/2-0.04$, Eq.~(\ref{eq:RSEoffset}) yields
\begin{eqnarray}
\Delta{L}_\mathrm{cd}+\Delta{L}_\mathrm{ctrl}\ (\zeta=0)=-\left(\frac{1+\rho}{1-\rho}\right)\frac{L\omega_\mathrm{c}\xi\tan{\phi}}{2\omega_0}\simeq-2\times10^{-9}~\mathrm{m}\ .
\end{eqnarray}
On one hand, having $\Delta L_\mathrm{cd}+\Delta L_\mathrm{ctrl}\gg \Delta L_\mathrm{rms}$ is often required to make the readout quadrature stable, on the other hand, $\Delta L_\mathrm{D}(=\Delta L_\mathrm{cd}+\Delta L_\mathrm{ctrl}+\Delta L_\mathrm{rms})$ should not be so big as to contribute significantly to laser noise. This will limit the range of available readout quadratures.

Figure~\ref{fig:RSEdcreadout} will help us to calculate laser noise in the detuned RSE configurations with DC readout. The reference light to probe the signal is remained carrier component that consists of contrast defect and the offset light excluding rms fluctuation. As is explained in Appendix~\ref{app:rms}, rms fluctuation is actually not a DC component but drifting at very low frequencies. The remained light should be bigger enough than rms fluctuation, otherwise the readout quadrature changes from time to time and causes problems in a data analysis of observed gravitational waves. Laser noise can be obtained from the calculation of the couplings between DC components and AC components of contrast defect and the offset light including rms fluctuation. Splitting $\mbox{\boldmath $b$}_\mathrm{RSE}$, shown in Eq.~\eqref{eq:RSE}, into its signal component and its noise component, one can write the output AC component of the detection with DC readout as
\begin{eqnarray}
\left(\mathrm{SG}+\mathrm{LN}\right)_{\rm RSE-DC}=\left(
\begin{array}{@{\,}c@{\,}}
B_1\\
B_2
\end{array}\right)_{\rm RSE-rem}
\bullet\left(
\begin{array}{@{\,}c@{\,}}
b_1\\
b_2
\end{array}\right)_{\rm RSE-sig}
+\left[\left(
\begin{array}{@{\,}c@{\,}}
B_1\\
B_2
\end{array}\right)_{\rm RSE-rem}
+\left(
\begin{array}{@{\,}c@{\,}}
B_1\\
B_2
\end{array}\right)_{\rm RSE-rms}\right]
\bullet\left(
\begin{array}{@{\,}c@{\,}}
b_1\\
b_2
\end{array}\right)_{\rm RSE-noise}\,.\label{eq:LNRSEDC}
\end{eqnarray}
Signal-referred laser noise is 
\begin{eqnarray}
h_\mathrm{rse-dc}&=&\frac{(D_{11}a_1+D_{12}a_2)\cos{\zeta}-(D_{21}a_1+D_{22}a_2)\sin{\zeta}}{D_{12}\cos{\zeta}-D_{22}\sin{\zeta}} \frac{\omega_{\rm c} -i\omega}{\sqrt{2}{\omega_0}}\nonumber\\
&&\ \ \ \ \ -\frac{(1+\rho)\sin{\phi}(D_{11}a_1+D_{12}a_2)-(1-\rho)\cos{\phi}(D_{21}a_1+D_{22}a_2)}{D_{12}\cos{\zeta}-D_{22}\sin{\zeta}} \frac{\sqrt{2}(\omega_{\rm c} -i\omega)\Delta L_\mathrm{rms}}{M_\mathrm{dc}\omega_\mathrm{c}L\xi'}\,.\label{eq:hrseDC}
\end{eqnarray}
The second term should be small and negligible unless we set $\Delta L_\mathrm{ctrl}=-\Delta L_\mathrm{cd}$ and the readout phase to be same as the quadrature of contrast defect light.

It is very important to say that we should not use $\zeta$ that is shown in Eq.~\eqref{eq:hrseDC} as the readout phase to calculate quantum noise using the definition of reference~\cite{Alessandra} since we have changed the definition in order to use the same equations for laser noise. The true readout phase in detuned RSE configuration with the DC readout scheme is 
\begin{eqnarray}
\zeta_\mathrm{DC}=\zeta-\frac{\pi}{2}\ ,
\end{eqnarray}
and Eq.~\eqref{eq:hrseDC} is rewritten as
\begin{eqnarray}
h_\mathrm{rse-dc}&=&\frac{(D_{11}a_1+D_{12}a_2)\sin{\zeta_\mathrm{DC}}+(D_{21}a_1+D_{22}a_2)\cos{\zeta_\mathrm{DC}}}{D_{12}\sin{\zeta_\mathrm{DC}}+D_{22}\cos{\zeta_\mathrm{DC}}} \frac{\omega_{\rm c} -i\omega}{\sqrt{2}{\omega_0}}\nonumber\\
&&\ \ \ \ \ +\frac{(1+\rho)\cos{\phi}(D_{11}a_1+D_{12}a_2)+(1-\rho)\sin{\phi}(D_{21}a_1+D_{22}a_2)}{D_{12}\sin{\zeta_\mathrm{DC}}+D_{22}\cos{\zeta_\mathrm{DC}}} \frac{\sqrt{2}(\omega_{\rm c} -i\omega)\Delta L_\mathrm{rms}}{M_\mathrm{dc}\omega_\mathrm{c}L\xi'}\,.\label{eq:hrseDCtrue}
\end{eqnarray}

Here we shall comment on the fundamental property of rms fluctuation. So far, an rms fluctuation term has been treated like DC light such as a contrast defect component or a control offset, and laser noise coupled via rms fluctuation has been linearly summed with other components. Sometimes this can be not true since the rms fluctuation term is not DC light but sum of low-frequency components that will couple with laser noise at the frequencies around a measurement frequency. Rigorously speaking, contribution of laser noise via rms fluctuation would be square-summed if we integrate the output of the detector longer than the time-scale of the drift in order to accumulate the gravitational-wave signal (see Appendix~\ref{app:rms}). In fact it depends on the way of data analysis, and the difference would be negligible in the cases with DC readout since rms fluctuation should be sufficiently smaller than the offset light. In the following calculation, we use the linear-sum assumption as has been written in this paper.

\subsection{Stabilization to the quantum level}

We shall see how much we can stabilize $\delta\nu$ and $\delta P$ before the comparison of laser noise with the detector sensitivity. Intensity stabilization is done by feeding back the intensity fluctuation of picked-off light before the injection to the interferometer. The quantum limit can be improved with the amount of the light that, on the other hand, should be small not to shave too much power used for the interferometer and not to saturate the output of photodetectors. Here we assume 100~mW. Shot-noise-limited sensitivity to measure the intensity fluctuation of 100~mW light is $\sim 5.5\times10^{-9}$~(W/W$\sqrt{\mathrm{Hz}}$), which is derived from
\begin{eqnarray}
\frac{\delta P}{4P}\sqrt{\frac{I_\mathrm{pick}}{\hbar\omega_0}}=1\ (\mathrm{quantum\ fluctuation})\ .
\end{eqnarray}
Frequency of the laser is stabilized to a common-mode motion of the arm cavities, which is measured by the signal detected at the bright port. The common-mode signal includes both frequency noise and the common-mode motion at the same time, and the high-frequency component is fed back to the laser so that the frequency is stabilized up to the level of the common-mode motion. This cut-off frequency can be lower than the observation band of the gravitational-wave detector so that one may assume, for simplicity, the stability is limited by the common-mode motion at all the frequencies. A difference between the signal detected at the bright port and that detected at the dark port is that the carrier light reflected back to the bright port increases shot noise by the proportion to the RF sidebands used as the reference light. The quantum-noise-limited sensitivity of the common mode is 
\begin{eqnarray}
h_\mathrm{com}=\sqrt{\frac{4\hbar}{\kappa_\mathrm{com}m\omega^2L^2}}\left[\left(\kappa_\mathrm{com}\frac{1+s_\mathrm{c}}{1+s_\mathrm{cc}}\frac{\delta P}{4P}\right)^2\frac{I_\mathrm{BS}}{\hbar\omega_0}+\left(1+\kappa_\mathrm{com}^2\right)+\left(\frac{\mathrm{Amp[Carrier+junk\ light]}}{\mathrm{Amp[SB]}}\right)^2\right]^{1/2}\ ,
\end{eqnarray}
with common-mode radiation pressure noise of shot-noise-limited intensity noise included. Here $\kappa_\mathrm{com}$ is
\begin{eqnarray}
\kappa_\mathrm{com}&=&\frac{I_\mathrm{BS}/I_\mathrm{SQL}'\cdot 2\omega_\mathrm{cc}^4}{\omega^2(\omega^2+\omega_\mathrm{cc}^2)},\\
I_\mathrm{SQL}'&=&\frac{mL^2\omega_\mathrm{cc}^4}{4\omega_0}\ .
\end{eqnarray}
Thus, quantum-noise-limited frequency-fluctuation level is given by $4\omega/\pi\cdot h_\mathrm{com}/\sqrt{8\hbar/\kappa_\mathrm{com}m\omega^2L^2}$. Assuming the proportion of the reference light to the total light at the bright port to be $10~\%$, we will have the fluctuation level of $\sim3\times10^{-9}~\mathrm{Hz}/\sqrt{\mathrm{Hz}}$ at 100~Hz, for example.

\subsection{Comparison of laser noise and detector sensitivity}
\begin{figure}[htbp]
\begin{center}
 \includegraphics*[width=12cm]{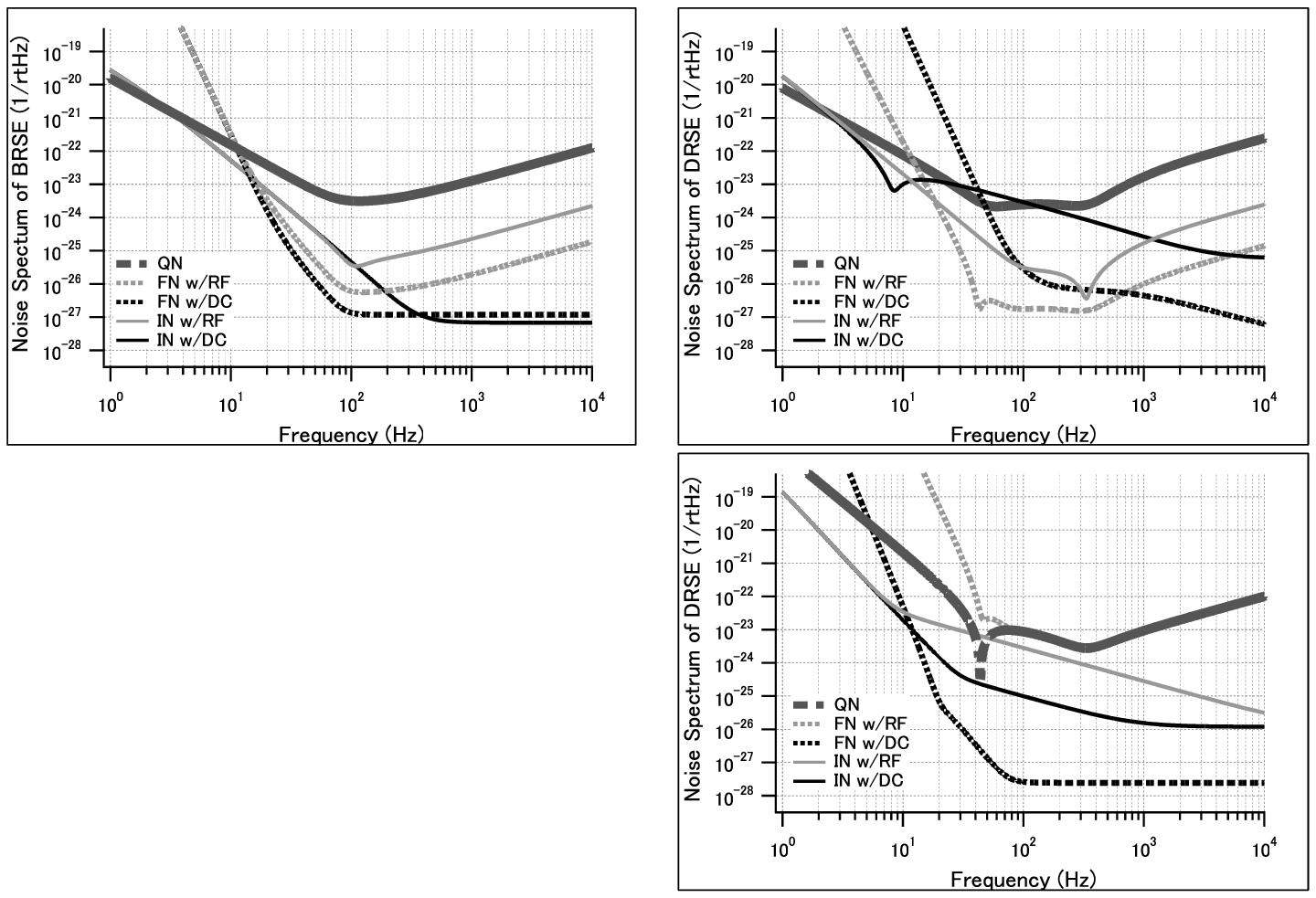}
 \caption{Quantum noise and laser noise of the broadband RSE (left), the detuned RSE with $\zeta\ \mathrm{or}\ \zeta_\mathrm{DC}=\pi/2$ (top-right), and the detuned RSE with $\zeta\ \mathrm{or}\ \zeta_\mathrm{DC}=0$ (bottom-right). Additional quantum noise at the demodulation process of RF readout is not included, which would possibly be removable in several ways~{\cite{AlessandraNSSN}\cite{SomiyaAspen}}.\label{fig:IgorSpectrum}}
\end{center}
\end{figure}

We have derived the transfer function of the light fluctuation to the dark port output and the fluctuation level of the light that is stabilized to the quantum noise level. Now we shall compare the laser noise level to the sensitivity of next generation detectors that should be nearly limited only by quantum noise. Figure~\ref{fig:IgorSpectrum} shows the results.

In the case of the broadband RSE, laser noise is mostly smaller than the quantum noise level. Frequency noise is larger than quantum noise at frequencies below $10\sim 20$~Hz, but seismic noise, which is not described here, will anyway limit the sensitivity at the frequencies. 

In the case of the detuned RSE, laser noise can limit the sensitivity if $\zeta$ is close to zero. This is due to the fact that the particular readout phase that gives the leaked carrier light at the dark port infinity is close to zero; see Eq.~(\ref{eq:particularphase}). Since the readout phases of RF readout ($\zeta$) and DC readout ($\zeta_\mathrm{DC}$) are different by $\pi/2$, laser noise with either scheme exceeds the level of quantum-noise-limited sensitivity and the other one does not in the observation band. Changing the readout phase between $0$ and $\pi/2$, we see laser noise start invading the sensitivity at around the optical-spring frequency when the readout phase becomes higher than $\sim\pi/2-0.004$ with DC readout, and lower than $\sim0.2$ with RF readout. The optical-spring dip is steep and the highest narrow-band sensitivity could be achieved at the frequency when the readout phase is close to zero. On the other hand, the sensitivity at low frequencies is better with the readout phase close to $\pi/2$. The readout phase should be chosen carefully with a consideration of laser noise due to our calculation and target gravitational-wave sources.

We have checked a consistency of the laser noise spectrum to the sign of the parameters used. Sometimes there appears or disappears a peak resulting from the cancellation of laser noise coupling via different paths, but the difference is trivial.

\section{Summary}
\label{sec:summary}
In this paper, we have provided  analytical calculations of laser noise in RSE interferometers with high optical power. We have done so by combining previous results of conventional interferometers, putting them into the quadrature representation, and then taking into account signal recycling.  Because of high optical power, laser noise can couple to the detection port not only optically, but also through driving mirror motions.  The latter effect can be dramatic at low frequencies. In particular, laser phase noise, coupled to the dark port via arm-length mismatch, will modulate carrier amplitude in the arms, after being re-injected back into the interferometer by the signal recycling mirror. As our result suggests, this noise can be  larger than quantum noise below $\sim$ 10 -- 20~Hz. Fortunately, this is nearly the edge of the observation band.

In the DC readout schemes, because we only have control of arm length, while contrast defect between the two arms provides a fixed output light, there is one quadrature which cannot be reached by adjusting differential control.  In order to achieve detection quadratures close to the forbidden one, a rather big arm-length mismatch will be needed, which will in turn increase laser noise. This eventually limits the accessible DC readout quadrature. 

In the RF readout scheme, one would also have DC light emerge at the orthogonal quadrature of the RF readout quadrature. If the required DC quadrature is close to the forbidden one, then the output DC light will be very high in amplitude,   and will increase laser noise. This constrains the RF readout phase.

\section*{Acknowledgement}
We wish to thank  Rana Adhikari, Osamu Miyakawa, Shigeo Nagano, Ken Strain, Alan Weinstein, and Hiro Yamamoto, for very useful discussions. We also wish to thank Thomas Corbitt for the open-use of his simulation software, with which most parts of our equations have been checked to be correct. This work is supported by the Japan Society for the Promotion of Science (JSPS) fellowship, the Sofja Kovalevskaja Programme (funded by the German Federal Ministry of Education and Research), and the LIGO Visitors Program.

\appendix
\section{RMS fluctuation}\label{app:rms}

In gravitational-wave interferometers, the microscopic arm length mismatch $\Delta L_\mathrm{D}$ will drift slowly at $\sim 1$~Hz, with a magnitude of $\sim 10^{-13}$~m  --- despite the seismic isolation system and low-frequency control system.  Such a magnitude may cause significant amount of laser noise to couple to the detection port; in DC readout schemes, it may also cause the readout quadrature, therefore the signal transfer function, to fluctuate significantly. Strictly speaking, we have a non-linear problem. The output of our detector will also have non-stationary response to gravitational waves, as well as non-stationary noise --- at a time scale of $\sim 1\,$second. Such non-stationarity will affect detections of different gravitational-wave sources in a different way.  To be most conservative, one has to
choose appropriate configurations and readout schemes, and impose the appropriate requirements on the level of drift, in such a way that the drift does not contribute significantly to the readout quadrature phase, nor to the laser-noise coupling.

Here we provide a  rough analysis on the effect of the drift, which will give us an idea of what it means for the drift component of $\Delta L_D$ to be small compared to other contributions.  First let us evaluate the noise spectrum, assuming an output with the form
\begin{equation}
N(t) =n_1(t)  + \eta_\mathrm{n}(t)n_2(t)\,,
\end{equation}
where $n_{1,2}(t)$ are stationary noises that originate from input laser noises, while $\eta_\mathrm{n}(t)$, which varies slowly, is induced by drifting. We assume all three random processes to have zero expectation value, and that $\eta_\mathrm{n}(t)$ is statistically independent from $n_{1,2}(t)$. The autocorrelation function of $N$ can be written as
\begin{equation}
C_N(t,t') = \left\langle \left[n_1(t)+\eta_\mathrm{n}(t)n_2(t)\right]\left[n_1(t')+\eta_\mathrm{n}(t')n_2(t')\right]\right\rangle = C_{n_1}(t-t') +  C_{\eta_\mathrm{n}}(t-t') C_{n_2}(t-t')\,.
\end{equation}
This means $N$ is stationary, with a spectrum of
\begin{equation}
S_N(f) = S_{n_1}(f) + \int df' S_{\eta_\mathrm{n}}(f') S_{n_2}(f-f')\,.
\end{equation}
If the noise spectrum of $n_2$ does not vary around $f$ at the scale of $\sim 1\,$Hz, we can then extract $S_{n_2}(f)$ out of the integral, and write
\begin{equation}
S_N(f) \approx  S_{n_1}(f) + \langle\eta_\mathrm{n}^2\rangle S_{n_2}(f)\,.
\end{equation}
This suggest that, as far as the laser-noise spectrum is concerned, we should replace the drift component of $\Delta L$ by $\Delta L_{\rm rms}$, and add it to other noise contributions {\it by quadrature}.
For the signal, we assume a signal output of
\begin{equation}
h(t) + \eta_\mathrm{s}(t) g(t)
\end{equation}
where $h(t)$ and $g(t)$ depend on the amplitude of the gravitational wave, and $\eta_\mathrm{s}(t)$ is a slowly varying component due to drifting, which has zero expectation value. Note that $\eta_\mathrm{s}(t)$ is correlated with $\eta_\mathrm{n}(t)$.  Suppose a template function $s(t)$ is applied to the signal output, then the signal contribution to the detection statistic is
\begin{equation}
\label{signalstatistic}
S=\int_{t_0}^{t_0+T} s(t) \left[h(t) + \eta_\mathrm{s}(t) g(t)\right]dt\,,
\end{equation}
where $t \in[t_0,t_0+T]$ is the duration of the possible signal. For small-duration signal (i.e., with $T \ll 1\,$second) with signal arrival rate far lower than 1\,Hz, the the signal component will depend on the instantaneous value of $\eta_\mathrm{s}(t_0)$, we could then write
\begin{equation}
S \sim \int_0^T s(t) \left[h(t) \pm \langle \sqrt{\eta_\mathrm{s}^2}\rangle g(t)\right]dt\,.
\end{equation}
For longer signals with signal arrival rate far lower than $1\,$Hz, we could use the above as a conservative estimate, because the integral in Eq.~\eqref{signalstatistic} now averages away some fluctuations in $\eta_\mathrm{s}(t)$.  This means the signal contribution should be calculated assuming the drift component of $\Delta L_D$ to take constant values within $(-\Delta L_{\rm rms},+\Delta L_{\rm rms})$, added {\it linearly} to other contributions.

\section{Downconversion of RF sidebands and laser noise around RF frequency}\label{app:RF}

In this section, we consider the down-conversion of laser noise around RF sidebands during demodulation process.  For the input laser, the optical field after RF phase modulation can be written as
\begin{equation}
\label{laserwithnoise}
\Re\left\{\left[1+\alpha(t) + i\psi(t) \right]\left[1+ \Gamma_{\rm in} e^{-i\omega_\mathrm{m} t} - \Gamma_{\rm in}e^{i\omega_\mathrm{m} t} \right]
E_0 e^{-i\omega_0 t} \right\}
\,,
\end{equation}
where $\alpha$ and $\psi$ are the input laser amplitude and phase noises in the time domain and $\Gamma_{\rm in}$ is the input RF modulation depth. At the detection port, we first consider only the RF sidebands and the laser-noise sidebands around them. Assuming that bright-port--dark-port transfer function does not depend on frequency at around $\omega_0 \pm \omega_\mathrm{m}$ (but the transfer function around $\omega_0+  \omega_\mathrm{m}$  may well differ from that around $\omega_0 - \omega_\mathrm{m}$), we can write the output RF fields as
\begin{eqnarray}
\label{outRF}
E_{\rm RF}(t) &=&\Re\left\{
\left[1+\alpha(t) + i\psi(t) \right]\left[\Gamma^+ e^{i\gamma^+}e^{-i\omega_\mathrm{m} t} + \Gamma^- e^{i\gamma^-} e^{i\omega_\mathrm{m} t} \right]
E_0 e^{-i\omega_0 t} \right\} \nonumber \\
&=&\left(\begin{array}{@{\,}c@{\,}}
\cos\omega_0 t \\
\sin\omega_0 t 
\end{array}\right)
\bullet \Big[
\underbrace{\Gamma^+ \mathbf{R}(\gamma^+ -\omega_\mathrm{m} t) }_{\mbox{upper sideband} \atop \mbox{modulation}}
+
\underbrace{\Gamma^- \mathbf{R}(\gamma^- +\omega_\mathrm{m} t)}_{\mbox{lower sideband} \atop \mbox{modulation} }\Big]\underbrace{E_0
\left[
\begin{array}{@{\,}c@{\,}}
1+ \alpha(t) \\ \psi(t)
\end{array}
\right]}_{\mbox{input laser}}\,.
\end{eqnarray}
Here $\Gamma^{+,-} \in \Re$ are upper and lower modulation depths at the detection port,  and $\gamma^{+,-} \in \Re$ represent phases of the sidebands; $\mathbf{R}$ is a rotation matrix, defined by
\begin{eqnarray}
\mathbf{R}(\theta) \equiv \left(
\begin{array}{cc}
\cos\theta & -\sin\theta \\
\sin\theta & \cos\theta
\end{array}
\right)
\end{eqnarray}
In the phasor diagram, the quadrature representation of Eq.~\eqref{outRF} (its second line)is very easy to understand: the input laser  with noises (last bracket) is being modulated to form an upper sideband and a lower sideband; they both rotate with frequency $\omega_\mathrm{m}$, but in opposite directions. Note that we have only considered laser noise ``inherited'' from the input laser,  i.e., upconverted to around the RF sidebands during phase modulation. Other fluctuations of the light is not considered, e.g., the intrinsic fluctuation of the input laser at around the RF frequency, which is not important.

On the other hand, output fields around the carrier can be written as
\begin{eqnarray}
E_{\rm carrier}(t)
=\left(\begin{array}{@{\,}c@{\,}}
\cos\omega_0 t \\
\sin\omega_0 t 
\end{array}\right)
\bullet
\left[
\begin{array}{@{\,}c@{\,}}
B_1 + b_1(t) \\
B_2 + b_2(t)
\end{array}
\right]_{\rm ca}\,.
\end{eqnarray}
The output photocurrent, given by $i(t) = \overline{2E_{\rm RF}(t)E_{\rm carrier}(t)}$, averaged over time scales longer than $1/\omega_0$, is
\begin{eqnarray}
\label{itheterodyne}
i(t)=
\left[\begin{array}{@{\,}c@{\,}}
B_1 + b_1(t)\\
B_2 + b_2(t)
\end{array}\right]_{\rm ca}
\bullet
\left[
\Gamma^+ \mathbf{R}(\gamma^+ -\omega_\mathrm{m} t)
+
\Gamma^- \mathbf{R}(\gamma^- +\omega_\mathrm{m} t)
\right]E_0
\left[
\begin{array}{@{\,}c@{\,}}
1+ \alpha(t) \\ \psi(t)
\end{array}
\right] \,.
\end{eqnarray}
During demodulation, $i(t)$ is mixed with $\sin(\omega_\mathrm{m} t +\delta_d)$, and then with its acoustic components extracted. This is equivalent to making the following substitution in Eq.~\eqref{itheterodyne},
\begin{eqnarray}
\label{demodintegral}
\left[
\Gamma^+ \mathbf{R}(\gamma^+ -\omega_\mathrm{m} t)
+
\Gamma^- \mathbf{R}(\gamma^- +\omega_\mathrm{m} t)
\right] &\rightarrow&
\frac{\omega_\mathrm{m}}{2\pi}\int_0^{2\pi/\omega_\mathrm{m}}   \left[
\Gamma^+ \mathbf{R}(\gamma^+ -\omega_\mathrm{m} t)
+
\Gamma^- \mathbf{R}(\gamma^- +\omega_\mathrm{m} t)\right]\sin(\omega_\mathrm{m} t +\delta_d)  dt
\nonumber \\
&=& \Gamma \left(
\begin{array}{cc}
\sin\zeta & -\cos\zeta \\
\cos\zeta & \sin\zeta
\end{array}
\right)\,.
\end{eqnarray}
This is possible because the integrand always has equal diagonal terms, and non-diagonal terms with opposite signs. Generically, varying the demodulation phase $\delta_d$ can give us any arbitrary $\zeta$. However, in special case of {\it balanced sidebands,} or $\Gamma^+ = \Gamma^-$, we will only be able to obtain $\zeta =(\pi-\gamma^+-\gamma^-)/2$, regardless of $\delta_d$.

Using Eq.~\eqref{demodintegral}, Eq.~\eqref{itheterodyne} can be written as
\begin{equation}
i(t) =
\left[\begin{array}{@{\,}c@{\,}}
B_1\\
B_2
\end{array}\right]_{\rm RF}
\bullet
\left[
\begin{array}{@{\,}c@{\,}}
b_1(t) \\ b_2(t)
\end{array}
\right]_{\rm ca} +
\left[\begin{array}{@{\,}c@{\,}}
B_1\\
B_2
\end{array}\right]_{\rm ca}
\bullet
\left[
\begin{array}{@{\,}c@{\,}}
b_1(t) \\ b_2(t)
\end{array}\right]_{\rm RF}
+ \mbox{(static or second-order terms)}
\,,
\end{equation}
with
\begin{equation}
\left[
\begin{array}{@{\,}c@{\,}}
B_1 \\  B_2
\end{array}
\right]_{\rm RF}
\equiv
\Gamma  E_0
\left(
\begin{array}{@{\,}c@{\,}}
\sin\zeta \\ \cos\zeta
\end{array}
\right)\,,\quad
\left[
\begin{array}{@{\,}c@{\,}}
b_1(t) \\  b_2(t)
\end{array}
\right]_{\rm RF}
\equiv
\Gamma E_0 \left(
\begin{array}{c	c}
\sin\zeta & -\cos\zeta \\
\cos\zeta & \sin\zeta
\end{array}
\right)
\left[
\begin{array}{@{\,}c@{\,}}
\alpha(t) \\ \psi(t)
\end{array}
\right]
\end{equation}
Here the quantities $\Gamma$ and $\zeta$ are not true modulation depth and modulation phase, but the result of weighted averaging done within the time interval of  $2\pi/\omega_\mathrm{m}$

\bibliographystyle{junsrt}

\begin{thebibliography}{99}
\bibitem{LIGO} D.~Shoemaker and The LIGO Scientific Collaboration, "Detector Description and Performance for the First Coincidence Observations between LIGO and GEO", Nucl. Inst. and Meth. in Phys. Res. A \textbf{517}, 154-179 (2004)
\bibitem{VIRGO} L.~Di~Fiore and VIRGO collaboration, "The present status of the VIRGO Central Interferometer," Class. Quant. Grav. \textbf{19}, 1421-1428 (2002)
\bibitem{GEO} B.~Willke {\it{et al}}, "The GEO 600 gravitational wave detector," Class. Quant. Grav. \textbf{19}, 1377-1387 (2002)
\bibitem{TAMA} M.~Ando and TAMA collaboration, "Stable Operation of a 300-m Laser Interferometer with Sufficient Sensitivity to Detect Gravitational-Wave Events within Our Galaxy," Phys. Rev. Let. \textbf{86} 3950-3954 (2001)
\bibitem{AdLIGO} D. Shoemaker, "Advanced LIGO: Context and Overview (Proposal to the NSF)," Technical report, \textbf{LIGO-M030023-00-M} (2003).
\bibitem{AdLIGO2} P.~Fritschel, "Second generation instruments for the Laser Interferometer Gravitational Wave Observatory (LIGO)," Proc. SPIE, 4856-39, 282 (2003)
\bibitem{LCGT} K.~Kuroda {\textit{et al}}, "Large-scale Cryogenic Gravitational wave Telescope," Int. J. Mod. Phy. D \textbf{8}, 557 (1999)
\bibitem{Mizuno} J.~Mizuno, "Comparison of optical configurations for laser-interferometric gravitational-wave detectors"(Ph.D. Thesis), Max-Planck-Institut f\"ur Quantenoptik, Germany, 1995
\bibitem{SomiyaAO} K.~Somiya {\textit{et al}}, "Development of a frequency-detuned interferometer as a prototype experiment for next-generation gravitational-wave detectors," Appl. Opt. \textbf{44}, 3179-3191 (2005)
\bibitem{SomiyaThesis} K.~Somiya, "Investigation of radiation pressure effect in a frequency-detuned interferometer and development of the readout scheme for a gravitational-wave detector"(Ph.D. Thesis), Univ. of Tokyo, Japan, 2004
\bibitem{Alessandra} A.~Buonanno and Y.~Chen, "Quantum noise in second generation, signal-recycled laser interferometric gravitational-wave detectors," Phys. Rev. D \textbf{64}, 042006 (2001)
\bibitem{Camp} J.~Camp, H.~Yamamoto, S.~Whitcomb, and D.~McClelland, "Analysis of light noise sources in a recycled Michelson interferometer with Fabry-Perot arms," J. Opt. Soc. Am. A \textbf{17}, 120 (2000)
\bibitem{Sigg} ISC group, D.~Sigg, ed., "Frequency Response of the LIGO interferometer," \textbf{LIGO-T970084-00 D} (1997)
\bibitem{JimThesis} J.~Mason, "Signal Extraction and Optical Design for an Advanced Gravitational Wave Interferometer"(Ph.D. Thesis), California Inst. of Tech., USA, 2001
\bibitem{Weiss} R.~Weiss, \textit{Quarterly Progress Report, MIT Research Lab of Electronics} \textbf{105} (1972) 54
\bibitem{DC} P.~Fritschel, talk at Technical Plenary Session of the LSC meeting 2003, Hannover, \textbf{LIGO-G030460-00 R} (2003)
\bibitem{Winkler} W.~Winkler has pointed out this in his internal document.
\bibitem{scaling} A.~Buonanno and Y.~Chen, Phys.~Rev.~{\bf D} 67,  062002 (2003).
\bibitem{SomiyaPRD} K.~Somiya, "New Photodetection Method Using Unbalanced Sidebands for Squeezed Quantum Noise in Gravitational Wave Interferometer," Phys. Rev. D \textbf{67} 122001 (2003)
\bibitem{KLMTV} H.~Kimble, Y.~Levin, A.~Matsko, K.~Thorne and S.~Vyatchanin, "Conversion of conventional gravitational-wave interferometers into quantum nondemolition interferometers by modifying their input and/or output optics," Phys. Rev. D \textbf{65}, 022002 (2002)
\bibitem{Somiya40m} K.~Somiya, "Analytical Calculation of Frequency Noise in the Spectrum of the 40m RSE Interferometer," \textbf{LIGO-T040180-00 R} (2004)
\bibitem{Miyakawa40m} O.~Miyakawa \textit{et al}, "Measurement of Optical Response of a Detuned Resonant Sideband Extraction Interferometer," \textbf{LIGO-P060007-00 R}, submitted
\bibitem{LCGT2} The parameters are to be optimized based upon the upcoming research of thermal noise at low temperature. The latest parameters, which has been updated just recently, are $r_\mathrm{s}=0.88$, ${\cal{F}}=1500$, and $g_\mathrm{pr}=11$: LCGT Collaboration, "Technical Report of LCGT," the report for the Technical Advisory Committee, Kashiwa (2005)
\bibitem{Rana} R.~Adhikari, talk at Technical Plenary Session of the LSC meeting 2005, Livingston, \textbf{LIGO-G050091-00 D} (2005)
\bibitem{AlessandraNSSN} A.~Buonanno, Y.~Chen, and N.~Mavalvala, "Quantum noise in laser-interferometer gravitational-wave detectors with a heterodyne readout scheme," Phys. Rev. D \textbf{67} 122005 (2003)
\bibitem{homodyne} S.~Vyatchanin and A.~Matsko, JETP \textbf{82}, 1007 (1996)
\bibitem{SomiyaAspen} K.~Somiya, talk at the Aspen Winter Conference on Gravitational Waves, Aspen, \textbf{LIGO-G040228-00 Z} (2004)

\end{thebibliography}

\end{document}